\documentclass[12pt, letterpaper]{article}

\pdfoutput=1

\usepackage{amsmath}
\usepackage{amsfonts}
\usepackage{amssymb}
\usepackage[colorlinks,citecolor=red,urlcolor=red,bookmarks=false,hypertexnames=true]{hyperref}

\usepackage{adjustbox}

\usepackage{cite}

\usepackage{tikz-feynman}

\usepackage{graphicx, physics, subcaption}
\usepackage{verbatim }

\numberwithin{equation}{section}

\usepackage{color}

\usepackage{amsmath, amssymb, graphics, epsfig, graphicx}
\usepackage{epsf}
\usepackage{epstopdf}
\usepackage {amssymb}
\newcommand{\nc}{\newcommand}
\nc{\ba}{\begin{eqnarray}}
\nc{\ea}{\end{eqnarray}}

\newcommand{\calR}{{\cal{R}}}
\newcommand{\calP}{{\cal{P}}}

\def\bfk{{\bf k}}
\def\bfq{{\bf q}}
\def\bfx{{\bf x}}

\nc{\cN}{ {\cal{N}} }

\usepackage[normalem]{ulem}

\begin{document}

\vspace{5mm}
\vspace{0.5cm}
\begin{center}

\def\thefootnote{\fnsymbol{footnote}}

{\bf\large Axion USR Inflation}
\\[0.5cm]

{ Alireza Talebian$\footnote{talebian@ipm.ir}$,   Hassan Firouzjahi$\footnote{firouz@ipm.ir}$ }   \\[0.5cm]
{\small \textit{ School of Astronomy, Institute for Research in Fundamental Sciences (IPM) \\ P.~O.~Box 19395-5746, Tehran, Iran}}

\end{center}

\vspace{.8cm}

\hrule \vspace{0.3cm}


\begin{abstract}

We consider a model of inflation in which the inflaton field is a rolling axion with a potential which is flat enough to support an intermediate phase of USR inflation. Because of  the Chern-Simons interaction,  one polarization of the gauge field experiences the tachyonic growth during the first slow-roll stage, inducing large corrections in curvature perturbations via the inverse decay effect. A non-trivial feature of our setup is that once the system enters the USR phase, the instability parameter falls off rapidly, terminating the gauge field production efficiently. Consequently, the power spectrum involves two separate peaks, the first peak is induced by the gauge field particles production 
 while the second peak is due to standard USR mechanism. We show that the power spectrum at the intermediate scales develops strong scale-dependence  $\propto k^m$ with the index $m >4$. 
Calculating  the bispectrum, we demonstrate that non-Gaussianities with non-trivial shapes and multiple peaks are generated in this setup.

\end{abstract}
\vspace{0.5cm} \hrule
\def\thefootnote{\arabic{footnote}}
\setcounter{footnote}{0}

\newpage
	
\section{Introduction}
\label{sec:intro}

Primordial inflation has become the dominant paradigm for the early Universe cosmology. It  provides not only a simple mechanism to resolve the conceptual shortcomings of the standard Big Bang cosmology but also it is well-supported by  Cosmic Microwave Background (CMB) and other cosmological observations \cite{Planck:2018jri, WMAP:2010qai}. Simplest models of inflation are based on the dynamics of a single scalar field, the inflaton field $\phi$, enjoying a nearly flat potential, supporting a long enough period of slow-roll (SR) inflation.  In standard scenarios of single field SR inflation, the amplitude of the local-type non-Gaussianity $f_{_{\rm NL}}$ is at the order of SR parameters \cite{Maldacena:2002vr} and is practically unobservable.

Ultra slow-roll (USR) is a simple single field setup with an exact flat potential \cite{Kinney:2005vj, Namjoo:2012aa, Martin:2012pe, Morse:2018kda, Dimopoulos:2017ged} during which the inflaton velocity falls off exponentially. Consequently, unlike the conventional SR setups, the curvature perturbations grows on superhorizon scales which provides a mechanism to violate the  Maldacena consistency condition
\cite{Maldacena:2002vr, Namjoo:2012aa, Chen:2013kta, Chen:2013aj, Chen:2013eea, Akhshik:2015rwa, Cai:2018dkf, Mooij:2015yka, Bravo:2017wyw, Pi:2022ysn}.  More specifically, the curvature perturbation power spectrum for modes which become superhorizon during USR phase scales like  ${\cal P}_{\cal R} \propto {\cal P}_{_{\rm CMB}} \, e^{6\Delta N}$ in which ${\cal P}_{_{\rm CMB}}$  is the CMB scale power spectrum and $\Delta N$ represents the  duration of USR phase. There have  been growing interests in the USR models  for Primordial Black Holes (PBHs) formation and to generate detectable Gravitational Waves (GWs) signals, \cite{Ivanov:1994pa, Garcia-Bellido:2017mdw, Germani:2017bcs, Biagetti:2018pjj, Firouzjahi:2023lzg}. For some references  on  PBHs from USR models and implications for the generation of GWs  see \cite{Khlopov:2008qy, Ozsoy:2023ryl, Byrnes:2021jka, Escriva:2022duf, Pi:2024jwt}.

Alternatively, the role of inflaton can be played by an axion field.  Axion is a pseudo Nambu-Goldstone Boson field \cite{Freese:1990rb, Adams:1992bn, Kim:2004rp, Dimopoulos:2005ac, Easther:2005zr, McAllister:2008hb, Kaloper:2008fb, Flauger:2009ab,  Arkani-Hamed:2003xts} which arises from the spontaneous breaking of Peccei-Quinn symmetry in the early Universe. 
It was originally introduced to solve the strong CP problem \cite{Peccei:1977hh, Weinberg:1977ma, Wilczek:1977pj, Preskill:1982cy, Abbott:1982af, Dine:1982ah, Sikivie:1983ip, Choi:2020rgn} and recently is considered as a well-motivated candidate for dark matter \cite{Hui:2016ltb, Adams:2022pbo}. The existence of pseudo-scalar fields  like axions during inflation can affect the inflationary dynamics.  In these models, the axion field $\phi$ is coupled to a $U(1)$ gauge field $A_\mu$ via the Chern-Simons type interaction,
\begin{equation}
	\label{int}
	\mathcal{L}_{\mathrm{int}} = - \frac{\tilde{\alpha}}{4}\, \frac{\phi}{M_{\rm Pl}} \, F^{\mu\nu}\tilde{F}_{\mu\nu} \,;
	\hspace{1cm}
	\tilde{\alpha} \equiv \frac{\alpha}{f_a}M_{\rm Pl} \, ,
\end{equation}
where ${\tilde F}^{ \mu \nu} = \frac{1}{2\sqrt{-g}}  \varepsilon^{\mu\nu\eta\sigma} F_{\eta\sigma}$ is the dual field of $F_{\mu\nu} = \partial_\mu A_\nu - \partial_\nu A_\mu$ with  $\varepsilon^{0123}=1$ and  
$M_{\rm Pl}$ is the reduced Planck mass.  The strength of the interaction is controlled by the dimensionless parameter $\tilde{\alpha}$ which is expressed in terms of the initial coupling constant $\alpha$ and the axion decay constant $f_a$.  The Lagrangian of \eqref{int} indicates a parity violating interaction causing interesting cosmological phenomena. As pointed out in Refs.~\cite{Anber:2006xt, Anber:2009ua, Anber:2012du, Barnaby:2010vf, Barnaby:2011vw, Barnaby:2011qe, Peloso:2016gqs}, the rolling of axion $\phi$  amplifies one polarization of the gauge field via the tachyonic instability.  The tachyonic growth of the gauge field quanta then backreacts on the axion field itself via the inverse decay effect, $A + A \rightarrow \delta\phi$, causing the enhancement of the scalar power spectra  and generation of  non-Gaussianity \cite{Anber:2006xt, Anber:2009ua, Anber:2012du, Barnaby:2010vf, Barnaby:2011vw, Barnaby:2011qe, Peloso:2016gqs,  Meerburg:2012id, Durrer:2024ibi}, formation of PBHs \cite{Linde:2012bt, Bugaev:2013fya, Talebian:2022jkb, Talebian:2022cwk}, generation of chiral GWs at the interferometer  and CMB scales 
\cite{Cook:2011hg,Barnaby:2011qe,Domcke:2016bkh, Sorbo:2011rz,Barnaby:2012xt,Namba:2015gja, Garcia-Bellido:2016dkw, Obata:2016xcr,Dimastrogiovanni:2016fuu, Dimastrogiovanni:2018xnn, Gorji:2020vnh, Salehian:2020dsf}  and production of primordial magnetic fields \cite{ Talebian:2020drj, Brandenburg:2024awd}.

The correction in power spectrum induced from the gauge field production 
takes the form  $\Delta {\cal P}_{\cal R}(k) \propto e^{4\pi\xi_k}$ \cite{Anber:2006xt, Anber:2009ua, Anber:2012du, Barnaby:2010vf, Barnaby:2011vw, Barnaby:2011qe, Peloso:2016gqs,  Meerburg:2012id, Durrer:2024ibi}
with numerical factors which will be specified later on. Here  $\xi$ is known as the ``\textit{instability parameter}'' defined as, 
\begin{align}
	\label{xi}
	\xi \equiv \abs{\tilde{\alpha}} \sqrt{\frac{\epsilon}{2}} \,;
	\hspace{1cm}
	\epsilon \equiv \frac{\dot{\phi}^2}{2M_{\rm Pl}^2H^2} \, ,
\end{align}
where $\epsilon$ is the first SR parameter and  $\xi_k$ refers to the value of $\xi$ when the mode $k$ leaves the horizon. During inflation, the parameter $\xi$ may be treated as a constant as its time variation is subleading in a SR expansion. The authors of \cite{Barnaby:2011vw} have also estimated the corrections in  the amplitude of non-Gaussianity induced by the gauge field production.

In this work, we study a setup of  inflation in which an axion field plays the role of inflaton and the potential $V(\phi)$ experiences an intermediate USR phase. More specifically, our setup is a three-stage SR-USR-SR model as follows. The first stage is a SR phase (SR1) during which the instability parameter reaches the regime $\xi \gtrsim 1$ and  there is a significant production of the  gauge field particles. It leads to the enhancement of curvature power spectrum as in 
the setups of \cite{Anber:2006xt, Anber:2009ua, Anber:2012du, Barnaby:2010vf, Barnaby:2011vw, Barnaby:2011qe, Peloso:2016gqs,  Meerburg:2012id, Durrer:2024ibi}. The second stage is a USR phase during which the rolling inflaton faces a strong Hubble friction and its velocity falls off exponentially. As a result, the instability parameter $\xi$ falls off exponentially, $\xi \propto \sqrt{\epsilon} \propto e^{-3\Delta N}$, and the generation of gauge field quanta stops effectively.  Hence, the gauge field production touches a maximum value at the end of SR1. On the other hand, during the USR phase, the curvature perturbation experiences a second growth, ${\calP_\calR} \propto \epsilon^{-1} \propto e^{6\Delta N}$. Finally, the USR phase is followed by the second SR stage (SR2) during which the system reaches its attractor phase and inflation ends followed by (p)reheating. As we shall see, there can be two peaks in power spectrum of the modes which leave the horizon during the intermediate period, one peak induced by the gauge field particle production and the second by the USR mechanism. However, the total power spectrum and the interplays of the peaks are far from the simple linear addition of the two effects. Similarly, the bispectrum  inherits complicated contributions from the axion particles production and the USR dynamics.

We require that the transition from SR1 to the USR stage and then from the USR phase to SR2 phase to be smooth. However, in order to follow the dynamics analytically, we consider an idealized situation in which the transition from SR1 to USR and then to SR2 are instantaneous. This simplification allows us to use the gluing techniques to match the solutions of curvature perturbations in the different stages. Having said this, even in an instantaneous transition, it will take some time for the system to relax to its attractor phase during the SR2 stage. This is controlled by a sharpness (relaxation) parameter \cite{Cai:2018dkf, Firouzjahi:2023aum, Firouzjahi:2023bkt} which plays important roles in our analysis as well.

The paper is organized as follows. In section \ref{sec-setup} we review the setup of axion inflation at the background level. In section \ref{sec:production_A} we study the production of gauge field quanta in our setup including a USR phase and then look at their backreaction effects on the inflationary background. In section \ref{sec:R} we calculate the total curvature perturbations arising from both vacuum fluctuations and the inverse decay mechanism. The analysis of two-point and three-point correlation functions of curvature perturbations 
are presented in sections \ref{Power}  and \ref{sub:3point} followed by summary and discussions in section \ref{sec:sum}.

	
\section{The Setup}
\label{sec-setup}

The setup we consider is given by the following action,
\begin{equation}\label{action}
S = \int {\rm d}^{4}x ~ \sqrt{-g} ~ \bigg[ \frac{M_{\rm Pl}^2\, }{2} R - \frac{1}{2}\partial_{\mu} \phi \partial^{\mu} \phi
- V(\phi) - \dfrac{1}{4}F_{\mu \nu}F^{\mu \nu} 
+ \mathcal{L}_{\mathrm{int}} \bigg] \,,
\end{equation}
in which  $R$ is the Ricci scalar and ${\cal L}_{\rm int}$ is the interaction Lagrangian given in Eq.  \eqref{int}. The potential $V(\phi)$ is such that to yield the three-stage SR-USR-SR inflation. More specifically, the potential is flat during the limited range $\phi_e< \phi<\phi_i$ to support the USR phase in which $\phi_i$ and $\phi_e$ are the starting and end points of USR phase in the field space respectively.   

We assume the gauge field has no background value so  the background geometry is given by an isotropic and spatially flat, Friedmann-Lemaitre-Robertson-Walker (FLRW) universe, with the metric
\begin{align}
{\rm d}s^2 = -{\rm d}t^2+a^2(t)~\delta_{ij}~{\rm d}x^i{\rm d}x^j \, ,
\end{align}
in which $a(t)$ is the scale factor and $t$ is the cosmic time.

Because of the Chern-Simons interaction, the gauge field perturbations $A_\mu = (A_0, \vec{A})$ are produced non-perturbatively \cite{Anber:2009ua} which can affect  the background evolution of the scale factor and the axion field. The  electric field $\vec E$ and the magnetic field $\vec B$ associated to the vector potential $\vec{A}$ are given via, 
\begin{align}
	\label{E-B}
	\vec E = - \frac{1}{a^2}  \partial_\tau\vec {A}\, , \quad \quad
	\vec B = \frac{1}{a^2} {\vec \nabla} \times \vec A \, ,
\end{align}
where $\tau$ is the conformal time defined via 
${\rm d} \tau \equiv {\rm d}t/a(t)$. 

The gauge field perturbations induce electromagnetic sources  so the background fields equations are modified as follows, 
\begin{align}
\label{KG-eq}
\ddot \phi + 3 H \dot \phi  +  V_\phi &=  \frac{\tilde{\alpha}}{M_{\rm Pl}}\langle  \vec E \cdot \vec B \rangle \,,
\\
\label{Friedmann-eq}
3 M_{\rm Pl}^2 H^2 - V(\phi)- \frac{1}{2} \dot \phi^2 &= \frac{1}{2}\langle   \vec E^2 + \vec B^2  \rangle \,,
\end{align}
where the dots denote derivatives with respect to the cosmic time $t$, $\langle \cal O \rangle$ denotes the quantum expectation value\footnote{Here, the mean field approximation is assumed in order to construct a homogeneous background from the amplified gauge field fluctuations. For equivalent approaches like stochastic or ensemble averages see \cite{Talebian:2019opf, Talebian:2020drj, Talebian:2021dfq, Talebian:2022jkb}.} for the operator $ \cal O $, and $H = \dot a(t)/a(t)$ is the Hubble expansion rate.  

The gauge field quanta  are produced by the homogeneous rolling of the axion field $\phi(\tau)$. Ignoring the inflaton and metric perturbations (see section 5 of \cite{Barnaby:2011vw} for the complete treatment), the equation of motion for $\vec A$ in Coulomb-radiation gauge
$A_0= \nabla \cdot \vec A=0$ is given by, 
\ba
\label{A-eq}
\Big(
\partial_\tau^2 -\nabla^2  - \frac{\tilde{\alpha}}{M_{\rm Pl}} \partial_\tau \phi ~{\vec \nabla} \times \Big) \vec A  =0 \,.
\ea

As mentioned previously, our setup is  a three-stage model of inflation in which a USR phase has been sandwiched between two phases of SR inflation $({\rm SR} \rightarrow {\rm USR} \rightarrow {\rm SR})$. During the first  SR stage, the SR parameter $\epsilon$ is nearly constant $\epsilon(\tau) \simeq \epsilon_i$,  but during the intermediate USR phase it falls off like $\tau^{6}$. Finally, after the USR phase $\epsilon(\tau)$ has a non-trivial dynamics but quickly it assumes its attractor value which is denoted by $\epsilon_f$. 
Correspondingly, we consider the following form of $\epsilon(\tau)$ during inflation,
\begin{align}
	\epsilon(\tau)  
	=
	\left\{
	\begin{array}{ll}
		\label{eq:epsilon_piecewise}
		\epsilon_i
		&
		\tau \leq \tau_i \hspace{2cm} \text{(SR1)
			}
		\\
		\\
		\epsilon_i  \big(\frac{\tau}{\tau_i} \big)^6  
		&
		\tau_i \leq \tau \leq \tau_e  \hspace{1.1cm} \text{(USR)}
		\\
		\\
		\epsilon_f 
		& 
		\tau > \tau_e  \hspace{2cm} \text{(attractor SR2)
			}
	\end{array} \, . 
	\right.
\end{align}
In this picture $\tau_i$ and $\tau_e$ represent the time of the start and the end of USR which are related to the number of e-fold of USR phase via 
$\Delta N = \ln(\tau_i/\tau_e)$. Correspondingly, the value of $\epsilon$ at the end of USR is given by $\epsilon_e= \epsilon_i e^{- 6 \Delta N}$. In Eq. (\ref{eq:epsilon_piecewise}), we have assumed the possibility that $\epsilon_f \neq \epsilon_e$. This is because  the inflationary system  may keep evolving after the USR phase before it reaches its attractor phase in the final SR phase. This is controlled by the sharpness (or relaxation) parameter $h$ defined via \cite{Cai:2018dkf, Firouzjahi:2023aum, Firouzjahi:2023bkt}, 
\ba
\label{h-def}
h \equiv \frac{6 \sqrt{2 \epsilon_f}}{\dot \phi(t_e)}= 
-6 \sqrt{\frac{\epsilon_f}{\epsilon_e} } \, .
\ea
Without loss of generality, we have assumed $\dot \phi <0$ so $h <0$. 
For a very sharp transition with $h \rightarrow -\infty$, the system reaches its attractor phase immediately after the USR phase. This is the limit which was studied in \cite{Namjoo:2012aa} yielding to $f_{_{\rm NL}}=5/2$. On the other hand, for a mild transition with $|h| \ll 1$, the relaxation period to reach the attractor phase is long and as shown in \cite{Cai:2018dkf}, much of non-Gaussianity generated during the USR phase is washed out. In the analysis below, in order to safely ignore the possible small SR corrections and to treat the system analytically, we consider sharp transitions with $|h| >1$. 

\begin{figure}[t!]
	\centering
	\includegraphics[width=0.49\linewidth]{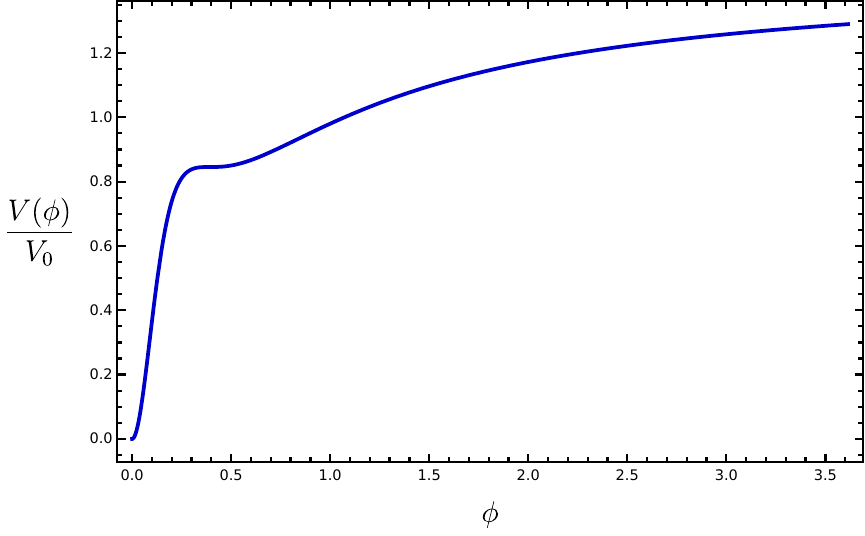}
	\includegraphics[width=0.48\linewidth]{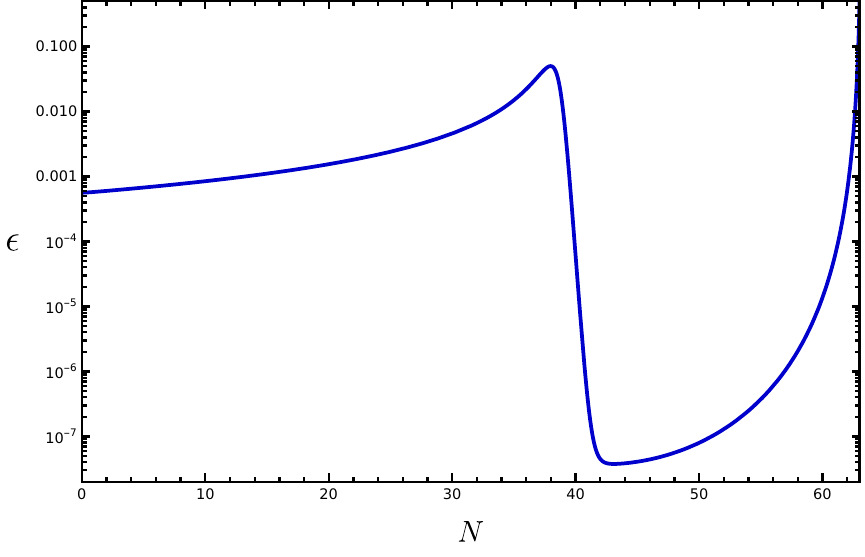}
	\caption{Left: The scalar potential $V(\phi)$ \eqref{potential}.  The parameters used are: $V_0 = 4 \times 10^{-10} M_{\rm Pl}^4, \nu = \sqrt{0.108} M_{\rm Pl}$ and $\lambda = 1.4349$. For these choices of parameters, the inflection point occurs at $\phi = 0.39 M_{\rm Pl}$. Right:  The behaviour of  $\epsilon(N)$  with the initial field value $\phi_i = 3.614 M_{\rm Pl}$. For these parameters the USR phase is sandwiched in the range $39-41$ e-folds and inflation lasts for about 63 e-folds.}
	\label{fig:epsilon_V}
\end{figure}

The discussions above were  mostly kinematical. To support the above picture dynamically, one needs potential with appropriate properties. The potentials supporting the above three-stage setup were proposed for example in  \cite{Ragavendra:2020sop, Taoso:2021uvl, Cheng:2021lif, Iacconi:2023ggt,Yang:2024ntt,Franciolini:2023agm}. As an example, and in order to follow the full numerical analysis,  we consider  the following toy potential  \cite{Garcia-Bellido:2017mdw, Ragavendra:2020sop}, 
\begin{align}
	\label{potential}
	V(\phi) = V_0 \frac{6x^2-4x^3+3x^4}{(1+\lambda ~x^2)^2} \,;
	\hspace{1cm}
	x=\frac{\phi}{\nu} \,,
\end{align}
where $\{V_0, \lambda,\nu\}$ are constant parameters of the model. This type of potential  arises  in models of Higgs Inflation \cite{Bezrukov:2007ep,Garcia-Bellido:2008ycs} which is asymptotically flat for large values of $\phi$ in order to be consistent with the CMB observations.  In Fig. \ref{fig:epsilon_V} the shape of the potential $V(\phi)$ and the evolution of $\epsilon(N)$ in term of number of e-fold ${\rm d}N\equiv H {\rm d}t$ are plotted with the parameters that allow for  about 63 e-folds of inflation. In this example, we have assumed that the field starts at larger values, well above the inflection point, and then slow-rolls down towards the minimum of the potential after crossing the intermediate USR stage.

\section{Production of Gauge Field Fluctuations}
\label{sec:production_A}
We consider the configuration that the gauge field has no classical background value $\langle A_\mu \rangle=0$ so the background equation (\ref{KG-eq}) 
has no source at the early times. However, as the gauge field are excited, 
they become tachyonic and their accumulated backreactions must be considered.

The gauge fields perturbations in Fourier space take the following form, 
\ba
\label{A_decomposition}
\vec A (\tau,x)=\sum_\lambda\int\frac{\dd[3]{k}}{(2\pi)^{3/2}} \vec \varepsilon_\lambda({\bf k})
\bigg[A_{\lambda}(k,\tau)~a_{{\lambda}}(\vb{k})
+A^*_{\lambda}(k,\tau)~a^\dagger_{{\lambda}}(-\vb{k}) \bigg] e^{i\vb{k}.\vb{x}}\,,
\ea
where $\vec \varepsilon_\lambda$ is the polarization vector with $\lambda =\pm$,  while $a_\lambda({\bf k})$ and $a_\lambda^\dagger({\bf k})$ are annihilation and creation operators satisfying
\begin{align}
	\label{a_a+}
	\left[a_\lambda({\bf k}), a_{\lambda'}^\dagger({\bf k'})\right] =\delta_{\lambda \lambda'} \delta^{(3)}({\bf k}-{\bf k'}) \,.
\end{align}
The polarization vectors satisfy the following 
conditions \cite{Barnaby:2011vw}, 
\begin{align}
	\label{pol-cond}
	\bfk \cdot \vec \varepsilon_\lambda(\bfk)=0  \, , \quad
	i {\bfk} \times \vec \varepsilon_\lambda(\bfk)= \lambda \, k \, \vec \varepsilon_\lambda(\bfk) \, ,\quad
	\vec \varepsilon_\lambda(-\bfk) = \vec \varepsilon_\lambda(\bfk)^* \, ,\quad
	\vec \varepsilon_\lambda(\bfk)^*\cdot \vec \varepsilon_{\lambda'}(\bfk) = \delta_{\lambda \lambda'} \, .
\end{align}

Plugging the decomposition \eqref{A_decomposition} into Eq. (\ref{A-eq}) yields the following equation for the vector field perturbations
\begin{align}
	\partial_\tau^2 A_\lambda+\Big[k^2-\lambda \big(\dfrac{\tilde{\alpha}}{M_{\rm Pl}}\dfrac{\dot{\phi}}{H}\big)\dfrac{k}{|\tau|}\Big]A_\lambda = 0 \,.
\end{align}
This equation shows that the wave-numbers with $k< \left|\frac{\tilde{\alpha}}{M_{\rm Pl}}\frac{\dot{\phi}}{H\tau}\right|$, corresponding to the polarization for which the combination $\lambda \, {\rm sign}(\dot{\phi}) \alpha$ is positive,
experience tachyonic instability (growing mode)  while the other polarization is damped. Hereafter, we denote the growing mode (decaying mode) by $A_+$ 
($A_-$), regardless of the actual sign of the polarization $\lambda$. 

\begin{figure}[t!]
	\centering
	\includegraphics[width=0.7\linewidth]{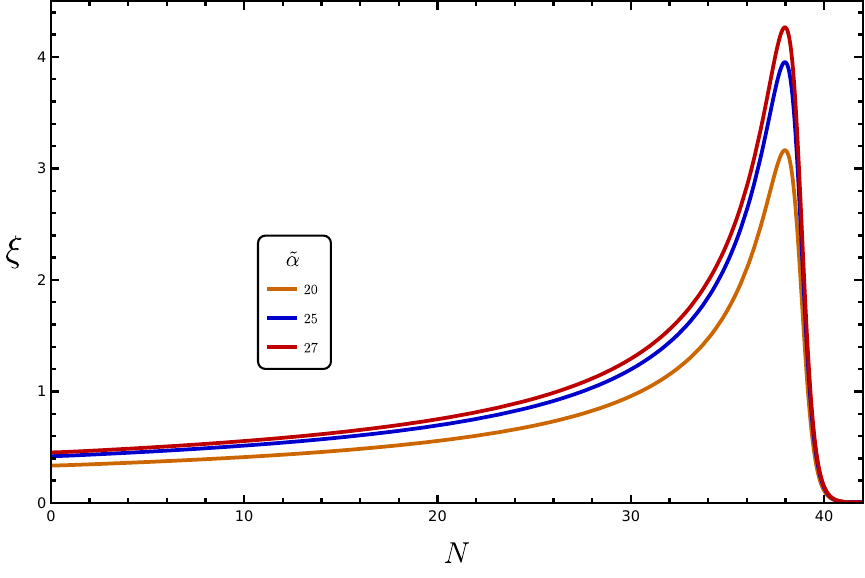}
	\caption{The behaviour of the instability parameter $\xi$  for various values of the coupling $\tilde{\alpha}$: $\tilde{\alpha}=20, 25$ and 27 with $\epsilon(N)$ shown as in  Fig. \ref{fig:epsilon_V}. The growth of  $\xi$ is terminated  effectively at the onset of the USR phase at around $N\simeq 39$.}
	\label{fig:xi}
\end{figure}

The dynamics of the growing mode is given by, 
\ba
\label{Aprime}
A_+''(z) + \Big( 1- \dfrac{2}{z}\xi(z) \Big) A_+(z) =0 \, ,
\ea
where the prime denotes the derivative with respect to $z \equiv -k\tau$ with the  instability parameter $\xi$ being defined in Eq. \eqref{xi}. Due to the non-trivial behaviour of $\epsilon(N)$ Eq.  \eqref{eq:epsilon_piecewise} during the three stages of inflation, the evolution of $\xi$ and consequently $A_+$ are different during these inflationary stages.  In Fig. \ref{fig:xi}, we have 
presented the behaviour of $\xi$  for various values of the coupling $\frac{\alpha}{f_a}$ and with $\epsilon$ plotted in Fig. \ref{fig:epsilon_V}. 
It is clearly seen that  the instability parameter falls off rapidly during the  USR phase and the tachyonic growth of the gauge field is turned off efficiently. 

In what follows, we solve the differential equation of \eqref{Aprime} for each phase separately and then match the solutions to obtain the final expression for the gauge fields.

During the first SR regime, the instability parameter $\xi$ is nearly constant, and the solution of \eqref{Aprime} can be expressed in terms of the Whittaker functions $W_{\mu,\nu}(z)$ and $M_{\mu,\nu}(z)$ as follows, 
\begin{align}
	A_+^{(1)}(z) = \dfrac{e^{\pi \xi_k /2}}{\sqrt{2k}} \Big(
	c_1 \, W_{-i\xi_k,\frac{1}{2}}(-2iz) + d_1 \, M_{-i\xi_k,\frac{1}{2}}(-2iz)
	\Big) \,,
\end{align}
in which $\xi_k$ is the instability parameter associated to the mode $k$ when it leaves the horizon which is nearly a constant. By using the asymptotic relation $W_{\mu,\nu}(z) \rightarrow z^\mu e^{-z/2}$ and $M_{\mu,\nu}(z) \rightarrow z^{-\mu} e^{z/2}$ for $z \rightarrow \infty$, the mode function $A(z)$ approaches the standard Bunch-Davies solutions $\frac{e^{iz}}{\sqrt{2k}}$ at early times $z \rightarrow \infty$  with $c_1=1$ and $d_1=0$. Correspondingly, the mode function is given by,  
\begin{align}
	\label{A_1}
	A_+^{(1)}(z) = \dfrac{e^{\pi \xi_k /2}}{\sqrt{2k}} W_{-i\xi_k,\frac{1}{2}}(-2iz) \,.
\end{align}

On the other hand, during the USR phase, $\epsilon(\tau) = \epsilon_i \big( \dfrac{\tau}{\tau_i} \big)^6$ and  the equation of motion \eqref{Aprime} is rewritten as, 
\begin{align}
	A_+''(z) + \Big( 1 - \Omega^2 z^2 \Big) A_+(z) =0 \, ;
	\hspace{1cm}
	\Omega^2 \equiv \frac{2|\tilde{\alpha}|}{(-k \tau_i)^3}\,\sqrt{\dfrac{\epsilon_i}{2}} \,. 
\end{align}
Correspondingly,  the solutions are given by,
\begin{align}
	A_+^{(2)}(z) = \dfrac{1}{\sqrt{z}} \Big(
	c_2 \, W_{\frac{1}{4\Omega},\frac{1}{4}}(\Omega z^2) + d_2 \,  M_{\frac{1}{4\Omega},\frac{1}{4}}(\Omega z^2) \Big) \, .
	\label{A_2}
\end{align}

During the final SR phase, after when the system has reached to its attractor phase, the evolution of gauge field is the same as \eqref{A_1} 
but with $\xi_k$  related to $\epsilon_f$. Considering both positive and negative frequency modes, the solution is given by, 
\begin{align}
	\label{A_3}
	A_+^{(3)}(z) = \dfrac{e^{\pi \xi_k /2}}{\sqrt{2k}} \bigg(
	c_3 \, W_{-i\xi_k,\frac{1}{2}}(-2iz) + d_3 \, M_{-i\xi_k,\frac{1}{2}}(-2iz)
	\bigg) \,,
\end{align}
with the coefficients $c_3$ and $d_3$ to be determined. 

Demanding the continuity of the gauge mode $A_+$ and its derivative $A_+'$ at $z_i = -k \tau_i$ and $z_e=-k\tau_e$, and  using the Wronskian relation for Whittaker functions\footnote{The Wronskian relation is
	\begin{align}
		{\cal W}\big\{W_{\mu,\nu}(y) , M_{\mu,\nu}(y) \big\}_y 
= \dfrac{\Gamma\left(1+2\nu\right)}{\Gamma \left(\frac{1}{2}+\nu-\mu \right)} \,.
		\end{align}
},
one can find the coefficients $c_{2,3}$ and $d_{2,3}$. 
We do not report the solutions of $c_{2,3}$ and $d_{2,3}$ here as they are 
too complicated to be useful but we use them in our analytical results in generating the numerical plots.

\subsection{Backreaction Effects}
\label{subsec:backreaction}

The tachyonic growth of the  gauge field perturbations affects  the dynamics of the  background fields through the sources in Eqs. \eqref{KG-eq} and \eqref{Friedmann-eq}. Consequently,  one distinguishes two distinct kinds of backreactions induced by gauge field perturbations.  First, the gauge field quanta are generated at the expense of the kinetic energy of the homogeneous axion $\phi(t)$. It introduces a new source of dissipation, an additional friction, into the right hand side of the inflaton equation of motion \eqref{KG-eq}. The second kind of the backreaction effect arises because the gauge field energy density contributes to the total energy density according to the Friedmann equation \eqref{Friedmann-eq}. We should examine that the rapid growth of gauge field fluctuations does not alter the background inflationary dynamics.

For this purpose, we define the  dimensionless parameters, 
\begin{align}
	\label{backreactions}
	\Omega_{\rm em} \equiv \dfrac{\langle
		\vec E^2 + \vec B^2
		 \rangle}{6M_{\rm Pl}^2H^2} \,,
	\hspace{1cm}
	S_{\rm em} \equiv \bigg| \dfrac{\tilde{\alpha} \langle \vec E \cdot \vec B \rangle}{3M_{\rm Pl}H\dot{\phi}} \bigg| \, ,
\end{align}
and investigate the following backreaction conditions, 
\begin{align}
	\label{healthy}
	\Omega_{\rm em} \ll 1  \,,
	\hspace{1cm}
	S_{\rm em} \ll 1 \, ,
\end{align}
in order to have a consistent inflationary scenario. 

\begin{figure}[t!!]
	\centering
	\includegraphics[width=0.2\linewidth]{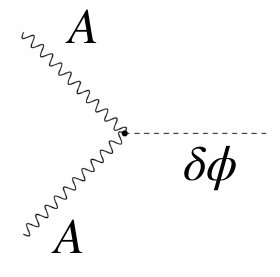}
	\caption{In the presence of the coupling $\phi F \tilde{F}$, gauge field quanta are copiously produced by the rolling axion. The produced gauge quanta in turn source inflaton fluctuations $\delta\phi$ via the inverse decay process  $A + A \rightarrow \delta \phi$  \cite{Barnaby:2011vw}. Dashed line denotes $\delta \phi$, while wiggly lines denote 
the gauge field $A$. 
}
	\label{fig:feynman}
\end{figure}

Even when the above two conditions are satisfied,  in the presence of the coupling $\phi F \tilde{F}$, 
the produced gauge quanta source inflaton fluctuations $\delta\phi$ via the  ``inverse decay'' process  $A + A \rightarrow \delta \phi$ \cite{Barnaby:2011vw}. Diagrammatic representation of the inverse decay process is shown in Fig. \ref{fig:feynman}. In section~\ref{sec:Inverse_Decay}, we 
calculate the contribution of this process in the primordial curvature perturbation.

\begin{figure}[t!]
	\centering
	\includegraphics[width=0.48\linewidth]{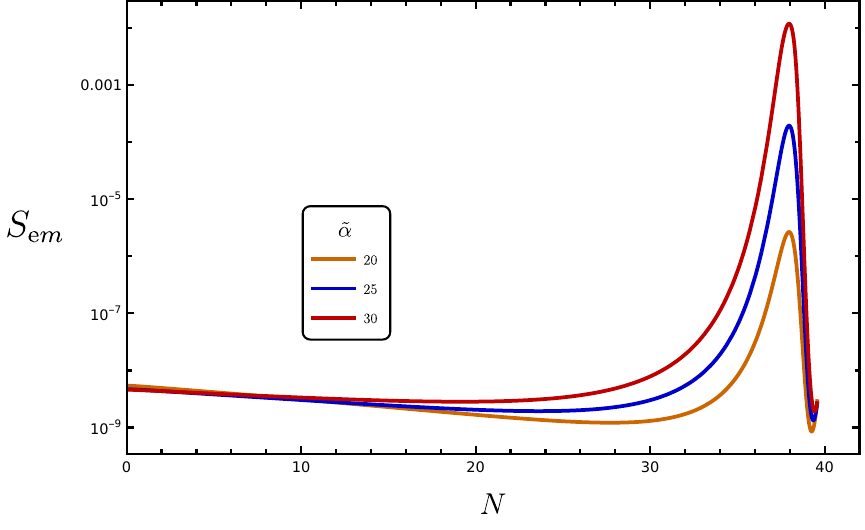}
	\includegraphics[width=0.45\linewidth]{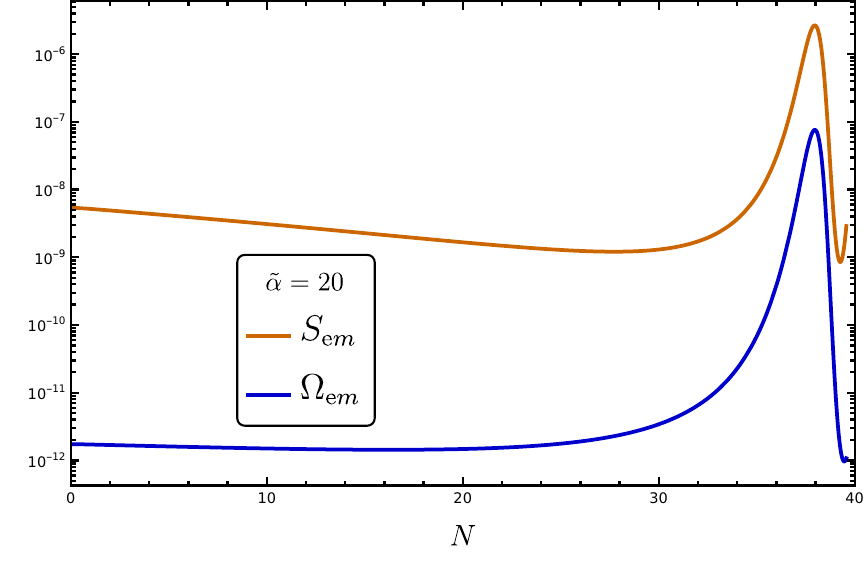}
	\caption{The behaviours of the backreaction parameters $S_{\rm em}$ and $\Omega_{\rm em}$. These plots are generated with the same parameters as used in Fig. \ref{fig:epsilon_V} but with various values of coupling $\tilde \alpha$. We see that while $S_{\rm em}$ and $\Omega_{\rm em}$ assume their maximum values at the end of SR1 phase, but the backreaction effects are negligible and they can be safely ignored.}
	\label{fig:back}
\end{figure}

From the structure of our setup, the largest value of $\xi$ occurs at 
the end of SR1 phase $\tau= \tau_i$, so the backreaction constraints by \eqref{backreactions} are most likely to be violated at $\tau= \tau_i$.  Therefore, we  evaluate them at this critical point. To this end, we have plotted the evolution of the quantities $S_{\rm em}$ and $\Omega_{\rm em}$ in Fig. \ref{fig:back}. These plots are generated  for parameters which were used to generate Fig. \ref{fig:epsilon_V} but with various values of the  coupling. 
These plots  demonstrate clearly  that the backreaction effects are negligible and can be safely ignored. In addition, Fig. \ref{fig:back} shows that the 
parameter  $S_{\rm em}$, controlling the backreaction on the inflaton dynamics in KG equation,  experiences more rapid growth than the second backreaction  parameter $\Omega_{\rm em}$. This is inline with the general conclusion that during the SR phase  the condition $S_{\rm em} \ll 1$ is more stringent than the condition $\Omega_{\rm em} \ll 1$ \cite{Talebian:2019opf, Talebian:2020drj}.

As seen in Fig. \ref{fig:xi}, the growth  of $\xi$ stops when the USR phase starts and before the backreaction term becomes significant. 
In conventional axion setup with an extended SR phase  and with a large coupling, the system experiences the nonlinear strong backreaction regime~\cite{Caravano:2022epk}. However,  thanks to the USR phase, our setup  does not enter into  this regime. Furthermore, $\xi$ does not experience the oscillatory epoch discussed in ~\cite{Cheng:2015oqa,Domcke:2020zez,Caravano:2022epk, Peloso:2022ovc}, again because before entering this phase at the end of SR1 stage, the USR dynamics terminates the growth of $\xi$. We work in the negligible backreaction regime such that the system never enter the oscillatory phase.


\section{Curvature Perturbations}
\label{sec:R}

At linear order in perturbations, the relation between the inflaton perturbations $\delta \phi$ and the curvature perturbation on the comoving hypersurfaces is given by,
\begin{align}
	\label{R}
	\calR\left( \tau ,\, \bfx \right) \equiv -\frac{H}{\dot \phi} \, \delta \phi\left( \tau ,\, \bfx \right) =  \frac{\delta \phi\left( \tau ,\, \bfx \right)}{M_{\rm Pl}\sqrt{2\epsilon(\tau)}} \, .
\end{align}
There are two new effects  in our setup compared to the usual inflationary models. The first effect is that during the USR phase  $\epsilon(\tau)$ evolves non-trivially according to Eq. \eqref{eq:epsilon_piecewise}. More specifically,  during the USR phase $\epsilon$ falls off exponentially so $\calR$ grows like $a(\tau)^3$. The second effect is the inverse decay process which induce a new contribution in 
 the inflaton perturbations $\delta \phi$. The produced gauge quanta source the inflaton fluctuations  which in turn contributes directly into the curvature perturbation \cite{Barnaby:2011vw}.

The evolution of the  inflaton fluctuations $\delta\phi$ can be cast in the following form \cite{Barnaby:2011vw},
\begin{align}
	\label{phi_eqn}
	\left[ \partial_\tau^2 +2 H \partial_\tau - \nabla^2 + a^2 V_{,\phi\phi} \right] \delta\phi(\tau,{\bf x}) 
	\simeq a^2 ~\frac{\alpha}{f_a}\Big(
	\vec E \cdot \vec B -\langle \vec E \cdot \vec B \rangle
	\Big) \,.
\end{align}
For weak backreaction regime  Eq. \eqref{phi_eqn} is trusted to a good approximation,  for more detail discussions see \cite{Anber:2009ua,Linde:2012bt}.

The solution of Eq. (\ref{phi_eqn}) can be decomposed into two parts: the homogeneous solution and the particular solution which is due to the source term.   Schematically, we have
\begin{equation}
	\delta\phi
	= \underbrace{\delta\phi^{(\mathrm{vac})}}_{\mathrm{homogeneous}} + \underbrace{\delta\phi^{(J)}}_{\mathrm{particular}} \,.
	\label{eq:vac_J}
\end{equation}
The superscripts ${(\rm vac)}$ and ${(J)}$ represent  the contribution from the   usual inflaton vacuum fluctuations  and the inverse decay processes  respectively. Both terms contribute to  curvature perturbation via  Eq.  \eqref{R}.

To proceed further, we decompose the inflaton fluctuations \eqref{eq:vac_J} in Fourier space. The homogeneous solution is expanded in terms of inflaton's ladder operators as follows, 
\begin{align}
	\label{delta_vac}
	a(\tau) \delta\phi^{\rm (vac)}(\tau,{\bf x}) &= \int \frac{\dd^3k}{(2\pi)^{3/2}}  \Big[
	{v_\bfk(\tau)}~b(\vb{k})
	+{v^*_{\bfk}(\tau)}~b^\dagger(-\vb{k})
	\Big] e^{i {\bf k}\cdot {\bf x}} \,,
\end{align}
where the annihilation/creation operators $b({\bf k}),b^\dagger({\bf k})$ associated with the inflaton vacuum fluctuations satisfy the standard commutation relation, 
\begin{align}
	\left[b({\bf k}), b^\dagger({\bf k'})\right] = \delta^{(3)}({\bf k}-{\bf k'})
	\label{b_b+} \, .
\end{align}

The particular solution can be expanded in terms of the  gauge field perturbations  as follows, 
\begin{align}
	\label{delta_J}
	a(\tau) \delta\phi^{(J)}(\tau,{\bf x}) &= \int \frac{\dd^3k}{(2\pi)^{3/2}}  
	{u_{\bf k}(\tau)} e^{i {\bf k}\cdot {\bf x}} \,,
\end{align}
where $u_{\bf k}$ includes the annihilation $a_\lambda({\bf k})$ and creation $a_\lambda^\dagger({\bf k})$ operators associated to the gauge field.  Note that at the free level, the inflaton and gauge field perturbations are independent quantum operators which commute with each other, 
\begin{equation}
	\label{ab_ladder}
	\left[b({\bf k}), a_\lambda({\bf k'})\right] = \left[b({\bf k}), a^\dagger_\lambda({\bf k'})\right] =0 \,.
\end{equation}
This means that the particular solution $\delta \phi^{(J)}$ is statistically independent of the homogeneous solution $\delta \phi^{\rm (vac)}$.  

Considering the small mass axion field during inflation ($V_{\phi\phi} \ll H^2$), the inflaton mode function $v_\bfk(\tau)$ is given by, 
\begin{equation}
	\label{eq:vac_mode}
	{v_\bfk(\tau)}\simeq 
	\dfrac{H  {a(\tau)} }{\sqrt{2k^3}} \bigg[
	\alpha_\bfk (1+ik\tau)e^{-ik\tau} + \beta_\bfk (1-ik\tau)e^{ik\tau}
	\bigg] \,, 
\end{equation}
in which the coefficients $\alpha_\bfk$ and $\beta_\bfk$ to be determined from the initial conditions. The special case of  Bunch-Davies initial condition when all modes are initially deep inside the horizon corresponds to  $\alpha_\bfk=1$ and $\beta_\bfk=0$. In addition, the particular mode function $u_k(\tau)$ can be expressed in terms of the retarded Green function as follows, 
\begin{align}
	\label{mode_u}
	u_{\bf k}(\tau) &= \int_{-\infty}^{0} \dd\tau' ~ G_k(\tau,\tau') ~J_{\bf k}(\tau'),
	\\
	\label{eq:Green}
	G_k(\tau,\tau') &= i \Theta(\tau-\tau') \Big[ v_k(\tau)v_k^*(\tau') - v_k^*(\tau)v_k(\tau') \Big] \, ,
\end{align}
where the source term is given by
\begin{align}
	\label{J}
	J_{k}(\tau) \equiv a^3(\tau)\frac{\tilde{\alpha}}{M_{\rm Pl}} \, 
	\int 
	\frac{\dd^3q}{(2\pi)^{3/2}} \, 
	{\vec E}_{\bf q}(\tau) \cdot {\vec B}_{{\bf k}-{\bf q}}(\tau) \,,
\end{align}
in which  ${\vec E}_{\bf q}(\tau)$ and ${\vec B}_{\bf q}(\tau)$ are the Fourier transforms of the electric and magnetic fields defined in real space in 
Eq. \eqref{E-B}.

Similar to Eq. \eqref{eq:vac_J}, the mode function of curvature perturbation $\calR_{\bfk}$ consists of two components: one associated to the usual vacuum fluctuations and the other arising from the gauge field inverse decay effect, written schematically as, 
\begin{equation}
	\calR_\bfk = \calR_\bfk^{(\mathrm{vac})} + \calR_\bfk^{(J)}
	\label{eq:R_vac_J} \, .
\end{equation}


\subsection{Curvature perturbations from vacuum fluctuations}

Here we calculate the first contribution in Eq. (\ref{eq:R_vac_J}), curvature perturbations generated from the inflationary vacuum fluctuations  
$\calR_\bfk^{(\mathrm{vac})}$. 

As mentioned before, to follow the dynamics analytically, we assume an idealized situation in which the transition from the SR1 to USR and then to final SR2 phases are instantaneous. This idealized picture can be relaxed by considering a more realistic situation  in which the transitions take some time. However, this modification will make the dynamics difficult to be solved analytically and a full numerical analysis is required. 

The comoving curvature perturbations at each phase is given by, 
\begin{align}
	\calR_k^{(\mathrm{vac})}(\tau) = - \frac{v_k(\tau)}{a(\tau)M_{\rm Pl}\sqrt{2\epsilon(\tau)}} \,,
\end{align}
where the mode function $v_k(\tau)$ is given by \eqref{eq:vac_mode}. 
With  $\epsilon(\tau)$ given in Eq.  \eqref{eq:epsilon_piecewise}, the curvature perturbation takes the following form, 
\begin{align}
	\calR_k^{(\mathrm{vac})}(\tau)  
	=
	\left\{
	\begin{array}{ll}
		\label{eq:R_piecewise}
		\calR_k^{(1)}(\tau) 
		&
		\tau \leq \tau_i \hspace{2cm} \text{( SR1 phase)}
		\\
		\\
		\calR_k^{(2)}(\tau) \propto \tau^{-3}
		&
		\tau_i \leq \tau \leq \tau_e  \hspace{1.1cm} \text{(USR phase)}
		\\
		\\
		\calR_k^{(3)}(\tau) 
		& 
		 \tau  > \tau_e \hspace{2cm} \text{( SR2 phase)}
	\end{array} \, . 
	\right.
\end{align}
In the following, we investigate the evolution of curvature perturbation in each phase.

\begin{enumerate}
	\item \textbf{First SR Phase $(\tau < \tau_i )$:}\\
	
	The system follows an attractor phase and the dynamics of the axion is given by the usual SR dynamics. The large CMB scale mode leaves the horizon during this stage and the first and second SR parameter $\epsilon$ and $\eta$ are nearly constant and small.
	
During the early stage  of inflation, all perturbations are deep inside the horizon. Starting with a Bunch-Davies initial condition during this phase with  $\alpha_k=1$ and $\beta_k=0$ in Eq. \eqref{eq:vac_mode}, the mode function is given by 
	\begin{equation}
		\label{eq:v1}
		{v_k^{(1)}(\tau)} \simeq 
		\dfrac{H {a(\tau)} }{\sqrt{2k^3}} (1+ik\tau)e^{-ik\tau} \, , 
	\end{equation}
and,
	\begin{align}
		\label{R_1}
		{\cal R}^{(1)}_k(\tau \leq \tau_i) = 
		\dfrac{H}{M_{\rm Pl}\sqrt{4k^3  \epsilon_i }}(1+i k \tau)
		e^{-i k \tau} \, ,
	\end{align}
where the superscript (1) indicates the first SR phase and $\epsilon_i$ is the value of $\epsilon$ before the USR phase starts.	
		
\item \textbf{USR Phase $(\tau_i < \tau < \tau_e)$:}\\

The USR phase takes over with a very  flat potential $V(\phi)\simeq V_0$. This phase is extended in the interval $\tau_i < \tau < \tau_e$ with  $\Delta N = \ln(\tau_i/\tau_e) $. 
	
As the potential is very flat during this stage,  $\dot{\phi}(t) \propto a(t)^{-3}$ so 
$\epsilon$ falls off exponentially, $\epsilon^{(2)} \propto a(t)^{-6}$, while the second slow-roll parameter $\eta \equiv \frac{\dot{\epsilon}}{H\epsilon}$ is nearly fixed with $\eta \simeq -6$ which is the hallmark of the USR phase. 
The value of $\epsilon$ at the end of USR phase $\epsilon_e$ is related to its value at the start of USR phase  via $\epsilon_e=\epsilon_i\, e^{-6\Delta N}$.
	
During the USR phase, the mode function is formally given by the superposition of the positive and negative frequency modes as in \eqref{eq:vac_mode}, 
	\begin{align}
		\label{R_2}
		{\cal R}^{(2)}_k (\tau_i \leq \tau \leq \tau_e)= \dfrac{H}{M_{\rm Pl}\sqrt{4\epsilon_i k^3}}\left(
		\dfrac{\tau_i}{\tau}
		\right)^3 \left[
		\alpha_k^{(2)}(1+i k \tau)e^{-i k \tau} + \beta_k^{(2)}(1-i k \tau)e^{i k \tau} \right] \, ,
	\end{align}
	where the coefficients $\alpha_k^{(2)}$ and $\beta_k^{(2)}$ are obtained by imposing  the continuity of ${\cal R}$
	and $\dot{\calR}$ at the transition point $\tau=\tau_i$~\cite{Firouzjahi:2023aum}.

\item \textbf{Final SR Phase $(\tau > \tau_e )$:}\\
	
After a few e-folds of inflation in the intermediate USR phase, the system enters into its final SR2 phase with a more extended period of inflation. With the assumption of an instantaneous transition from USR to SR, $\epsilon(\tau)$ after the USR phase is given by \cite{Cai:2018dkf, Firouzjahi:2023lzg},
	\begin{align}
		\label{ep-eq}
		\epsilon^{(3)}(\tau) &= \epsilon_ e  \big(\frac{h}{6}\big)^2\bigg[ 1 - \big( \frac{6+h}{h} \big) \big(\frac{\tau}{\tau_e} \big)^3 \bigg]^{2}  \, ,
	\end{align}
where the sharpness parameter $h$ defined in Eq. (\ref{h-def}) controls 
how quickly the system reaches to its attractor phase.  In the limit $h \rightarrow -\infty$, the system reaches its final attractor phase immediately after $\tau_e$ and the mode functions is frozen. On the other hand, a particular case of transition is when $h=-6$ in which $\epsilon(\tau)$ is fixed to its value at the end of USR, $\epsilon^{(3)}(\tau) =\epsilon_e$ with $\eta(\tau)=0$ for $\tau> \tau_e$~ \cite{Firouzjahi:2023aum}.
	
During this stage, the mode function has the following  form,
	\begin{equation}
		\label{v_3}
		\dfrac{v_k^{(3)}(\tau)}{a(\tau)} \simeq 
		\dfrac{H }{\sqrt{2k^3}} \Big[ \alpha^{(3)}_k ( 1+ i k \tau) e^{- i k \tau}  + \beta^{(3)}_k ( 1- i k \tau) e^{ i k \tau}  \Big] \,, 
	\end{equation}
with the curvature perturbation in the third phase  given by,
	\ba
	\label{mode3}
	\label{R_3}
	\calR^{(3)}_{k} (\tau_e < \tau )=  \frac{H}{ M_{\rm Pl}\sqrt{4 \epsilon^{(3)}(\tau) k^3}}  
	\Big[ \alpha^{(3)}_k ( 1+ i k \tau) e^{- i k \tau}  + \beta^{(3)}_k ( 1- i k \tau) e^{ i k \tau}  \Big] \, , 
	\ea
with $\epsilon^{(3)}(\tau)$ given by Eq. (\ref{ep-eq}). The coefficients $\alpha^{(3)}_k$ and $\beta^{(3)}_k$, after imposing the matching conditions at $\tau=\tau_e$, are obtained to be \cite{Firouzjahi:2023aum}, 
	\begin{align}
		\label{alpha_3}
		\alpha^{(3)}_k(h,z_i,z_e) &= \frac{1}{8 
			z_i^3 z_e^3}  \Big[ 3h
		( 1 +i z_e)^2 (1-i z_i)^2
		e^{2i (z_i- z_e)}-i (2 z_i^3 + 3i  z_i^2 - 3 i) (4 i z_e^3+ h z_e^2 + h) \Big]  \,,
		\\
		\label{beta_3}
		\beta^{(3)}_k (h,z_i,z_e)&=   \frac{1}{8 z_i^3 z_e^3}  \Big[ 3 ( 1 - i z_i)^2 ( h+ h z_e^2 - 4 i z_e^3 ) e^{2 i z_i} 
		+ i h ( 1- i z_e)^2  ( 3 i + 3 i z_i^2 - 2 z_i^3 ) e^{2 i z_e}
		\Big] \, ,
	\end{align}
where $z_{i,e}\equiv -k\tau_{i,e}$. 	
\end{enumerate}

We are interested in the curvature perturbations at the end of inflation, $\tau \rightarrow 0^-$, which from  Eq. \eqref{mode3} is given by, 
\begin{align}
	\label{R_vac_0}
	\calR_k^{(\mathrm{vac})}(0) = \calR^{(3)}_{k}(0) = \frac{H}{ M_{\rm Pl}\sqrt{4  k^3\epsilon_f}}  
	\Big( \alpha^{(3)}_k   + \beta^{(3)}_k   \Big) \, .
\end{align}
The USR parameters $\{h, \tau_i, \tau_e\}$ affect the observed curvature perturbations via their impact on $\alpha^{(3)}_k$ and $\beta^{(3)}_k$.


\subsection{Curvature perturbations from inverse decay effect}
\label{sec:Inverse_Decay}

In this subsection we calculate the curvature perturbations induced by 
the generation of gauge field particles.

The correction in curvature perturbation from the inverse decay $\calR_k^{(J)}(\tau)$ is given by \cite{Barnaby:2011vw}, 
\begin{align}
	\calR_k^{(J)}(\tau) = - \frac{u_k(\tau)}{a(\tau)M_{\rm Pl}\sqrt{2\epsilon(\tau)}} \, ,
\end{align}
where the particular mode function $u_k(\tau)$ is given by the retarded Green function in Eqs. \eqref{eq:Green} and  \eqref{mode_u}. 
Correspondingly, at the time of end of inflation we obtain,
\begin{align}
	\label{R_J_0}
	\calR_k^{(J)}(0) = - \frac{i}{M_{\rm Pl}\sqrt{2\epsilon_f}}\int_{-\infty}^{0} \dd\tau' ~ \Big[ \frac{v_k(0)}{a(0)}v_k^*(\tau') - \frac{v_k^*(0)}{a(0)}v_k(\tau') \Big] ~J_{k}(\tau') \, ,
\end{align}
where  we have assumed the system has reached to its attractor phase with $\epsilon(\tau) \rightarrow \epsilon_f$. The source term $J_{k}(\tau)$, defined in Eq. \eqref{J}, is controlled by the electric and magnetic fields generated by the  rolling axion field.

As the rolling of inflaton is different during the three stages, the induced 
curvature perturbation from the gauge field production has three distinct forms. 

\begin{enumerate}
	\item \textbf{First SR phase:}
	Due to the slow-rolling of axion field,  one of the polarizations of $\vec A$ experiences a tachyonic instability with significant  gauge field particles production  as governed by Eq. \eqref{Aprime}. 
The produced gauge field backreacts on the  inflaton fluctuations via inverse decay, modifying the  curvature perturbation as, 
	\begin{align}
		{\cal R}_\bfk = {\cal R}^{(1)}_\bfk + {\cal R}^{(J)}_\bfk \,.
	\end{align}
In section~\ref{sec:Inverse_Decay} we  compute the contribution of ${\cal R}^{(J)}_\bfk$ into the power spectrum.
	
\item \textbf{USR phase:}
	
During this stage, the instability parameter $\xi$ falls off exponentially and tachyonic production of gauge field is effectively terminated. The produced gauge fields during the first SR phase is diluted like radiation. Therefore, at the background level, the dynamics of the system is given by \eqref{KG-eq} and \eqref{Friedmann-eq} without contribution from the right-hand sides. 
As the tachyonic instability of the gauge modes is terminated, the inverse decay process is  inefficient and the particular solution of \eqref{phi_eqn} can be safely ignored during this phase. Correspondingly, for the perturbations leaving the horizon during the USR phase, only the inflaton perturbations contribute into $\calR$ and  
\begin{align}
		{\cal R}_k \simeq  {\cal R}_k^{\mathrm{(vac)}} =  
		{\cal R}^{(2)}_k  \,.
\end{align}

\item \textbf{Final SR phase:}
During the final SR stage, the tachyonic mechanism is resumed which in principle can  result in gauge field particle  productions. However, the gauge field production during this phase is not as efficient as in the latest stage of the SR1 phase because  $\epsilon_f$ is typically much smaller 
then $\epsilon_i$ and the instability parameter is small, $\xi^{(3)} \ll 1$. Therefore, during this phase we have	
	\begin{align}
	{\cal R}_k \simeq {\cal R}_k^{\mathrm{(vac)}} =  {\cal R}^{(3)}_k  \,.
	\end{align}
\end{enumerate}

The conclusion from the above discussions is that  the gauge field is practically sourced during the first SR phase so  the source term can be parametrized as follows,
\begin{align}
	\label{J_theta}
	J_k(\tau') \propto \Theta(\tau_i-\tau') \,.
\end{align}
This in turn  affects the upper limit of integral \eqref{R_J_0}, yielding 
\begin{align}
	\label{R_J_0_2}
	\calR_k^{(J)}(0) = - \frac{i}{M_{\rm Pl}\sqrt{2\epsilon_f}}\int_{-\infty}^{\tau_i} \dd\tau' ~ \bigg[ \frac{v_k(0)}{a(0)}\frac{v_k^*(\tau')}{a(\tau')} - \frac{v_k^*(0)}{a(0)}\frac{v_k(\tau')}{a(\tau')} \bigg] ~a(\tau') J_{k}(\tau') \, ,
\end{align}
where $v_k(\tau')$ corresponds to the mode function during SR1 phase, given by \eqref{eq:v1}, and $v_k(0)$ corresponds to the mode function at the final SR2 phase as stated in \eqref{v_3}. This expression can be simplified as follows, 
\begin{align}
	\frac{v_k(0)}{a(0)}\frac{v_k^*(\tau')}{a(\tau')} - \frac{v_k^*(0)}{a(0)}\frac{v_k(\tau')}{a(\tau')} &= \frac{iH(0)H(\tau')}{k^3} \Im\Big[
	\big(
	\alpha^{(3)}_k+\beta^{(3)}_k
	\big)^* ( 1+ i k \tau) e^{- i k \tau} \Big] \, ,
\end{align}
where we have used the fact that during inflation $H(0) \simeq H(\tau')=H$.

\subsubsection{Source term $J_k(\tau)$}

To calculate the source term $J_k(\tau)$ in Eq. \eqref{J} we need to calculate the  electric and magnetic fields.  During the  SR1 phase  the gauge field evolves according to the solution in Eq. \eqref{A_1}. This solution was obtained from the assumption that  $\xi$ varies slowly, $\frac{\dot{\xi}}{\xi H} \ll 1$. For $\xi \gtrsim {\cal O}(1)$, the evolution of  the gauge field fluctuations can be approximated as
\begin{align}
	\label{A_+_approx}
	A_+(k,\tau < \tau_i) \simeq 
	\dfrac{e^{\pi \xi_k-2\sqrt{2z\xi_k}}}{\sqrt{2k}} \left(
	\frac{z}{2\xi_k}
	\right)^{1/4} \,;
	\hspace{1cm}
	z \equiv  -k \tau \, ,
\end{align}
where we  omit the superscript $(1)$ for $A_+^{(1)}$ in the rest of this paper.
This solution is for the modes satisfying $(8\xi_k)^{-1}\lesssim z \lesssim 2\xi_k$  which account for most of the power contained in the gauge field fluctuations\cite{Barnaby:2010vf}.  Since only the positive helicity modes 
with real values are amplified,  the gauge fields perturbation can be written as, 
\begin{align}
	\vec A(\tau, x) =\int \frac{{\rm d}^3 k}{(2\pi)^{3/2}}  ~e^{i\vb{k}.\vb{x}}~ \vec{\mathcal{O}}_{\bf{k}} ~A_+(k,\tau) \,,
\end{align}
with
\begin{align}
	\label{O}
	\vec{\mathcal{O}}_{\bfk} \equiv {\vec\varepsilon}_+(\bfk)\left[a_+(\bfk)+a_+^{\dagger}(-\bfk)\right] \,.
\end{align}
Correspondingly, the electric and magnetic fields are given by 
\begin{align}
	\vec E(\tau, x) &= \int \frac{{\rm d}^3 k}{(2\pi)^{3/2}} ~e^{i\vb{k}.\vb{x}}~{\vec E}_{\bf k}(\tau)  \,,
	\\
	\vec B(\tau, x) &=  \int \frac{{\rm d}^3 k}{(2\pi)^{3/2}}  ~~e^{i\vb{k}.\vb{x}}~{\vec B}_{\bf k}(\tau)  \,,
\end{align}
where the Fourier components appearing in the source term, are defined as
\begin{align}
	\vec E_{\bf k}(\tau) &= \frac{-1}{a^2(\tau)}~A_+'(k,\tau)~\vec{\mathcal{O}}_{\bf{k}} \,,
	\\
	\vec B_{\bf k}(\tau) &= \frac{\abs{\bfk}}{a^2(\tau)} ~A_+(k,\tau)~\vec{\mathcal{O}}_{\bf{k}} \, .
\end{align}

With the above solutions of the electric and magnetic fields, the source term  $J_k(\tau) $ in Eq. \eqref{J} can be approximated as
\begin{align}
	\label{J_approx}
	J_{k}(\tau) \simeq \left(
	\frac{-\tilde{\alpha}}{M_{\rm Pl}}
	\right)\frac{{e^{2\pi \xi_k}}}{4a(\tau)} 
	\int 
	\frac{\dd^3q}{(2\pi)^{3/2}} \, \abs{\bfk-\vb{q}}^{1/4}\abs{\vb{q}}^{1/4}
	~
	\left(
	\abs{\bfk-\vb{q}}^{1/2}+\abs{\vb{q}}^{1/2}
	\right)
	~
	\nonumber\\
	\times g(-\tau,\abs{\vb{q}},\abs{\bfk-\vb{q}})~
	\vec{\mathcal{O}}_{\bf{q}} \cdot \vec{\mathcal{O}}_{\bf{k}-\bf{q}}
	\,,
\end{align}
where \eqref{A_+_approx} is used for $A_+(k,\tau)$ and the function $g(x,\abs{\vb{q}},\abs{\bfk-\vb{q}})$ is defined as follows, 
\begin{align}
	g(x,\abs{\vb{q}},\abs{\bfk-\vb{q}}) \equiv e^{-2\sqrt{2x}\left(
		\abs{\vb{q}}^{1/2}+\abs{\bfk-\vb{q}}^{1/2}
		\right)} \, .
\end{align}

Putting all together,  the contribution of the gauge field perturbations through 
inverse decay effects in the curvature perturbation at the end of inflation is calculated to be, 
\begin{align}
	\label{R_J_00}
	\calR_k^{(J)}(0) = &
	\left(\frac{-\tilde{\alpha}}{M_{\rm Pl}}\right)
	\frac{H^2 e^{2\pi\xi_k}}{4k^3M_{\rm Pl}{\sqrt{2\epsilon_f}}}
	\int 
	\frac{\dd^3q}{(2\pi)^{3/2}}~\, \abs{\bfk-\vb{q}}^{1/4}\abs{\vb{q}}^{1/4} \left(
	\abs{\bfk-\vb{q}}^{1/2}+\abs{\vb{q}}^{1/2}
	\right) ~
	\vec{\mathcal{O}}_{\bf{q}} \cdot \vec{\mathcal{O}}_{\bf{k}-\bf{q}} \, 
	\nonumber\\
	&~~~~~ ~~~~~ \times \int_{-\infty}^{\tau_i} \dd\tau' ~ \Im\bigg[
	\left(
	\alpha^{(3)}_k+\beta^{(3)}_k
	\right)^* ( 1+ i k \tau) e^{- i k \tau}
	\bigg] 
	\, g(-\tau'\xi_k,\abs{\vb{q}},\abs{\bfk-\vb{q}}) \,.
\end{align}
Here   $\xi_k$ is the value of the instability parameter when the mode $k$ leaves the horizon during the SR1 stage so $\xi_k$ depends on the wave number $k$ implicitly.

Note that $\alpha^{(3)}_k$ and $\beta^{(3)}_k$ depend on the  USR parameters $\{h, \tau_i, \tau_e\}$. In the previous works of axion inflation 
involving a single SR phase such as in \cite{Barnaby:2011vw}, $\alpha^{(3)}_k=1$ and $\beta^{(3)}_k=0$. In addition, in comparison to these models  the upper bound of the integral in our model in Eq. (\ref{R_J_00}) is limited to $\tau_i$ and not to the end of inflation $\tau =0$. Finally, in comparison with the previous works, there is an additional modification due to the fact that $\epsilon_f \neq \epsilon_i$.\\

Finally, putting all together, the total curvature perturbation at the end of inflation is,
\begin{align}
	\label{R_0}
	\calR_\bfk(0) = \calR_\bfk^{(\mathrm{vac})}(0) +  \calR_\bfk^{(J)}(0) \, ,
\end{align}
where $\calR_\bfk^{(\mathrm{vac})}(0)$ and $\calR_\bfk^{(J)}(0)$ are given by Eqs. \eqref{R_vac_0} and \eqref{R_J_00}, respectively. In the next section, we calculate the power spectrum and bispectrum associated to  Eq. (\ref{R_0}).


\section{Power Spectrum}
\label{Power}

In this section we calculate the power spectrum. From Eq. (\ref{R_0}) 
both the usual vacuum fluctuations $\calR_\bfk^{(\mathrm{vac})}$ from 
the inflaton perturbations and $\calR_\bfk^{(J)}$  from the  gauge field particle production  contribute to the power spectrum.

Since $\calR_\bfk^{(\mathrm{vac})}$ and $\calR_\bfk^{(J)}$ are statistically independent, we have
\begin{align}
	\langle \calR_{\bf k} \,  \calR_{\bf k'} \rangle &=\langle \calR_{\bf k}^{(\mathrm{vac})} \,  \calR_{\bf k'}^{(\mathrm{vac})} \rangle + \langle \calR_{\bf k}^{(J)} \,  \calR_{\bf k'}^{(J)} \rangle \, .
\end{align}
Correspondingly,  we can decompose the power spectrum as follows, 
\begin{align}
	\label{eq:P_total}
	\calP_\calR (k,\tau) = \calP^{(\mathrm{vac})}_\calR (k,\tau) + \calP^{(J)}_\calR (k,\tau) \, .
\end{align}

\begin{figure}[t]
\vspace{-1cm}
	\centering
	\includegraphics[width=0.3\linewidth]{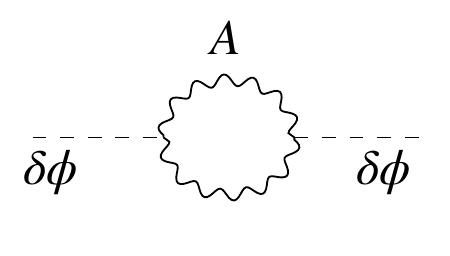}
	\caption{Diagrammatic representation of the scalar power spectrum sourced by the  gauge field perturbations.  
}
	\label{fig:loop}
\end{figure}

Physically speaking, the corrections from the  interaction $\phi F \tilde F$
into the  power spectrum,  $\calP^{(J)}_\calR (k,\tau)$, can be viewed as the one-loop correction represented in Fig. \ref{fig:loop}. As seen in this figure, we have two vertices involving the coupling $\tilde \alpha$ while the gauge field perturbations are propagating in the internal lines. However, note that this loop correction is totally different from the one-loop corrections in models of single field inflation  involving a USR phase for PBHs formation which are discussed in the recent literature, 
e.g. \cite{Kristiano:2022maq, Firouzjahi:2023aum}. In the latter setups, the loop corrections are induced from the scalar perturbations which are amplified during the USR phase while in our setup the loop corrections are generated by the tachyonic gauge field perturbations.

The power spectrum  is related to the two-point correlation function in the momentum space  via,
\begin{equation}
	\langle \calR_{\bf k}(\tau) ~ \calR_{\bf k'}(\tau) \rangle \equiv 
	\frac{2\pi^2}{k^3}
	\calP_\calR (k,\tau)  \,  \delta^{(3)}({\bf k}+{\bf k'}) \, .
	\label{power} 
\end{equation}

The physical quantities are calculated at the time of end of inflation $\tau=0$.
The contribution from the vacuum modes is standard, which from Eq. \eqref{R_vac_0} yields, 
\begin{align}
\label{P_vac}
\calP^{(\mathrm{vac})}_\calR(k) &= 
{\cal P}_{_{\rm CMB}} \big(\dfrac{\epsilon_i}{\epsilon_{_f}}\big)\big| \alpha^{(3)}_k   + \beta^{(3)}_k \big|^2  \, ,
\end{align}
where we have used the definition of the power spectrum at the CMB scales, 
\begin{align}
	\label{P_CMB}
	{\cal P}_{_{\rm CMB}} \equiv \dfrac{H^2}{8\pi^2M_{\rm Pl}^2\, \epsilon_{i}} \,.
\end{align}

On the other hand, the particular (source) solution \eqref{R_J_00}  contributes to the final power spectrum as follows, 
\begin{align}
	\label{P_J}
	\calP^{(J)}_\calR(k) &\simeq \frac{H^4}{ 2^{10} \pi^5M_{\rm Pl}^2 {\epsilon_f} } \bigg(\frac{\tilde{\alpha}}{M_{\rm Pl}}
	\bigg)^2e^{4\pi\xi_k} 
	\nonumber
	\\
	\times
	&\int \dd^3r
	\left[ 1 + \frac{\vert {\bf r} \vert^2 - {\bf r} \cdot {\hat \bfk}}{\vert {\bf r} \vert \vert {\hat \bfk} - {\bf r} \vert } \right]^2
	\vert {\bf r} \vert^{\frac{1}{2}} \vert {\bf r} - {\hat \bfk} \vert^{\frac{1}{2}} \left[ \vert {\bf r} \vert^{\frac{1}{2}} + \vert {\bf r} - {\hat \bfk} \vert^{\frac{1}{2}} \right]^2
	{\cal I}^2(\xi_k,\abs{\bf r},|\vb{r}-\hat{\bfk}|)  \, ,
\end{align}
where the following relations have been used, 
\begin{align}
	\langle \vec{\mathcal{O}}_{\bf{q}} \cdot \vec{\mathcal{O}}_{\bf{k}-\bf{q}} ~~ \vec{\mathcal{O}}_{\bf{q}'} \cdot \vec{\mathcal{O}}_{\bf{k}'-\bf{q}'} \rangle &= 8 \Big\vert {\vec\varepsilon}_+(\vb{q}) \cdot {\vec\varepsilon}_+({\bfk}-\vb{q}) \Big\vert^2 \, \delta^{(3)}(\bf{k}-\bf{q}+\bf{q}') \, \delta^{(3)}(\bf{k}'-\bf{q}'+\bf{q}) \,,\\ \abs{{\vec\varepsilon}_+(\vb{r}) \cdot {\vec\varepsilon}_+(\hat{\bfk}-{\bf r})}^2 &=\frac{1}{4} \left[ 1 + \frac{\vert {\bf r} \vert^2 - {\bf r} \cdot {\hat \bfk}}{\vert {\bf r} \vert \vert {\hat \bfk} - {\bf r} \vert } \right]^2 \,,
	\\
	{\vec\varepsilon}_+(-\vb{r}) &= {\vec\varepsilon}_+^{\,\, *}(\vb{r}) \,.	
\end{align}
In addition, we have defined the dimensionless integration variable ${\bf r}\equiv {\bf q}/\abs{\bf k}$, and introduced the following integral,
\begin{align}
	\label{eq:calI}
	{\cal I}(y,\abs{\bf r},|\vb{r}-\hat{\bfk}|) &\equiv \int_{z_i}^{\infty} \dd z ~\Im\Big[\big(\alpha^{(3)}_k   + \beta^{(3)}_k\big)^*(1-i z)e^{iz} \Big] ~ g(z y,\abs{\bf r},|\vb{r}-\hat{\bfk}|) \,.
\end{align}
Note that  the coefficients $\alpha^{(3)}_k$ and $\beta^{(3)}_k$ depend on the  parameters of  the  USR phase $\{h,z_i=-k\tau_i,z_e=-k\tau_e\}$. Considering the explicit forms \eqref{alpha_3} and \eqref{beta_3} for $\alpha^{(3)}_k$ and $\beta^{(3)}_k$, one can decompose the integral ${\cal I}$ as,
\begin{align}
	\label{eq:calI_1}
	{\cal I}
	&= {\cal F}_1(h,z_i,z_e)~{\cal I}_1+{\cal F}_2(h,z_i,z_e)~{\cal I}_2
	\,,
\end{align}
where 
\begin{align}
	\label{I1}
	{\cal I}_1(y,\abs{\bf r},|\vb{r}-\hat{\bfk}|)
	&= \int_{z_i}^{\infty} \dd z ~\bigg(\sin z-z\cos z\bigg)~ g(z y,\abs{\bf r},|\vb{r}-\hat{\bfk}|)
	\,,
	\\
	\label{I2}
	{\cal I}_2(y,\abs{\bf r},|\vb{r}-\hat{\bfk}|)
	&= \int_{z_i}^{\infty} \dd z ~\bigg(\cos z+z\sin z\bigg)~ g(zy,\abs{\bf r},|\vb{r}-\hat{\bfk}|)
	\,,
\end{align}
and ${\cal F}_{i=1,2}(h,z_i,z_e)$ are two given functions which we do not report here as they are too complicated to be insightful. 

In the absence of USR, e.g. the slow-roll axion model such as in \cite{Barnaby:2011vw}, we have $\alpha^{(3)}=1$ and $\beta^{(3)}=0$ yielding  to 
${\cal F}_{1}=1, {\cal F}_2=0$  with the lower bound of the integral \eqref{I1} is set to  zero.

Using  Eq. \eqref{P_CMB} for the power spectrum of CMB scales, 
we can parametrize the power spectrum induced by gauge field as, 
\begin{align}
	\label{P_J_2}
	\calP^{(J)}_\calR(k) ={\cal P}_{_{\rm CMB}}^2 \big(\dfrac{\epsilon_i}{\epsilon_{_f}}\big) f_2(\xi_k) ~ e^{4\pi\xi_k}  \, ,
\end{align}
where the function  $f_2(\xi_k) $ is defined as, 
\begin{align}
	\label{eq:f_2}
	f_2(\xi_k) \equiv& \dfrac{\xi_{_{\rm CMB}}^2}{8\pi} \int \dd^3r
	\Big( 1 + \frac{\vert {\bf r} \vert^2 - {\bf r} \cdot {\hat \bfk}}{\vert {\bf r} \vert \vert {\hat \bfk} - {\bf r} \vert } \Big)^2  \vert {\bf r} \vert^{\frac{1}{2}} \vert {\bf r} - {\hat \bfk} \vert^{\frac{1}{2}} \Big( \vert {\bf r} \vert^{\frac{1}{2}} + \vert {\bf r} - {\hat \bfk} \vert^{\frac{1}{2}} \Big)^2
	{\cal I}^2(\xi_k,\abs{\bf r},|\vb{r}-\hat{\bfk}|) \, .
\end{align}
In conventional SR axion models, this function depends only on $\xi$. In \cite{Barnaby:2011vw}, it has been estimated as,
\ba
\label{f2}
f_2(\xi) \simeq \frac{{\cal O} (10 ^{-5})}{\xi^6}  \, ,
\quad \quad ( \text{SR setup}) \, .
\ea 
However, in our axion USR setup, the parameters  of the USR phase $\{h,\tau_i,\tau_e\}$ contribute into ${\cal I}$ according to Eq. \eqref{eq:calI_1} so
$f_2(\xi_k)$ depends on these parameters as well.

Note that the dominant contribution of the integrals ${\cal I}_1$ \eqref{I1} and ${\cal I}_2$ \eqref{I2} comes from the lower end $z_i \rightarrow 0$. Therefore, it is expected that the value of function $f_2(\xi)$ in our case to be substantially smaller than the  value given in Eq. (\ref{f2}) for the conventional SR axion model.  It is convenient to define the effective function $f_2^{\rm eff}$ given below to compare its value  with the corresponding quantity in  conventional SR axion models, 
\begin{align}
	\label{eq:f2eff}
	f_2^{\rm eff}(\xi_k) &\equiv \dfrac{\epsilon_{i}}{\epsilon_{f}} f_2(\xi_k)
	\nonumber\\
	&=\Big(\dfrac{\xi_{_{\rm CMB}}}{\xi_i}\Big)^2 e^{6\Delta N} \Big(\dfrac{36}{h^2}\Big) f_2(\xi_k) \,,
\end{align}
where the following relations have been used, 
\begin{align}
	\dfrac{\epsilon_{_{i}}}{\epsilon_{f}} =    \dfrac{\epsilon_i}{\epsilon_e} \dfrac{\epsilon_e}{\epsilon_{f}}
	=\Big(\dfrac{\xi_{_{\rm CMB}}}{\xi_i}\Big)^2 e^{6\Delta N} \Big(\dfrac{36}{h^2}\Big) \,.
\end{align}
We see that the exponential enhancement $e^{6\Delta N}$ in \eqref{eq:f2eff} can compensate  the reduction in the value of $f_2(\xi_k)$ caused by the change in the lower bounds of integrals ${\cal I}_1$ and ${\cal I}_2$.
 
\begin{figure}[t]
	\centerline{
		\includegraphics[angle=0,width=0.7\textwidth]{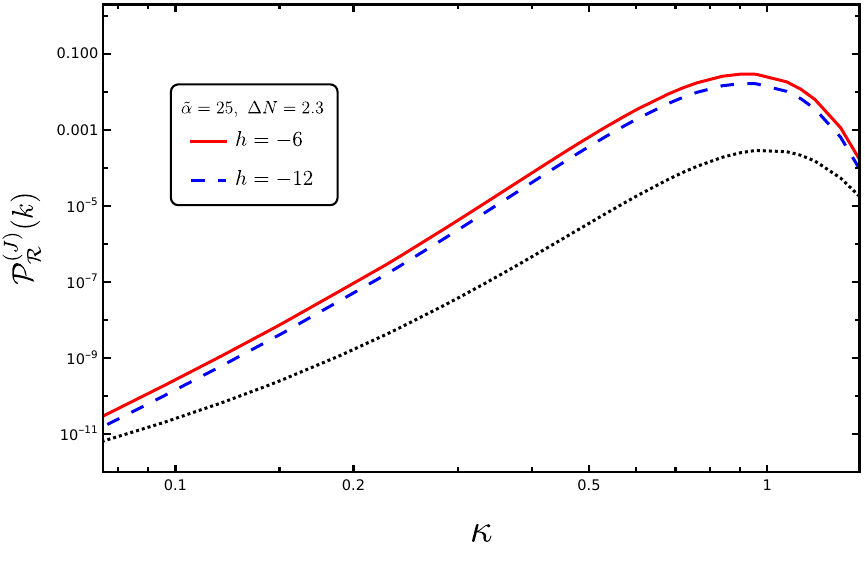}
	}
	\caption{The effects of the USR phase on the power spectrum 
	$\calP^{(J)}_\calR(k)$ \eqref{P_J_2} in terms of $\kappa \equiv \frac{k}{k_i}$ where $k_i$ represents the mode which leaves the horizon at the start of USR phase.  Here,  $\tilde{\alpha}=25$ while the USR parameters are
 $\Delta N=2.3$ and $h=\{-6,-12\}$. For comparison, the lower dotted black curve represents  the restricted case where USR  phase affects only the instability parameter, keeping the form of $f_2$ unchanged as given in Eq.  \eqref{f2}. We see that the exponential enhancement $e^{6\Delta N}$  largely offsets the reduction in $f_2(\xi_k)$ in \eqref{eq:f2eff} induced by the shift in the lower bounds of ${\cal I}_1$ and ${\cal I}_2$.
	}
	\label{fig:f2}
\end{figure}

In Fig. \ref{fig:f2}, we have plotted ${\calP^{(J)}_\calR}(k)$, the corrections in power spectrum induced  by the gauge field,  
for the modes which leave the horizon prior to the USR phase where the inverse decay effects develop its maximum. In comparison, we also presented  ${\calP^{(J)}_\calR}(k)$  in a modified scenario where the USR phase affects only the evolution of the   instability parameter, leaving the function $f_2$ unchanged as given in Eq. \eqref{f2} (unlike our case where USR modifies both).  As can be seen, the exponential factor $e^{6\Delta N}$ in \eqref{eq:f2eff}  largely offsets the suppression induced on $f_2(\xi_k)$ from the modification of the lower bounds of integrals ${\cal I}_1$ and ${\cal I}_2$. The conclusion is that the USR phase affects not only the vacuum curvature perturbations ${\calP^{\rm (vac)}_\calR}(k)$ but also ${\calP^{(J)}_\calR}(k)$.

Finally, adding \eqref{P_vac} and \eqref{P_J_2}, the total power spectrum of curvature perturbation \eqref{eq:P_total} is given by 
\begin{align}
	\label{eq:P_total_2_talk}
	\calP_\calR (k) &= {\cal P}_{_{\rm CMB}} \big(\dfrac{\xi_{_{\rm CMB}}}{\xi_i}\big)^2 e^{6\Delta N} \big(\dfrac{36}{h^2}\big)\Big(
	\big| \alpha^{(3)}_k   + \beta^{(3)}_k \big|^2 + {\cal P}_{_{\rm CMB}} f_2(\xi_k)e^{4\pi\xi_k} 
	\Big)
	\, .
\end{align}
The effects of the USR phase on ${\calP^{(J)}_\calR}(k)$ 
is encoded in the non-trivial function $f_2(\xi_k)$. In general, this function depends on the details of the USR phase through parameters $\{h, \tau_{{i}}, \tau_{{e}}\}$ as well as the wave number through $z=-k \tau$.

\begin{figure}[t]
	\centering
	\includegraphics[width=0.8\linewidth]{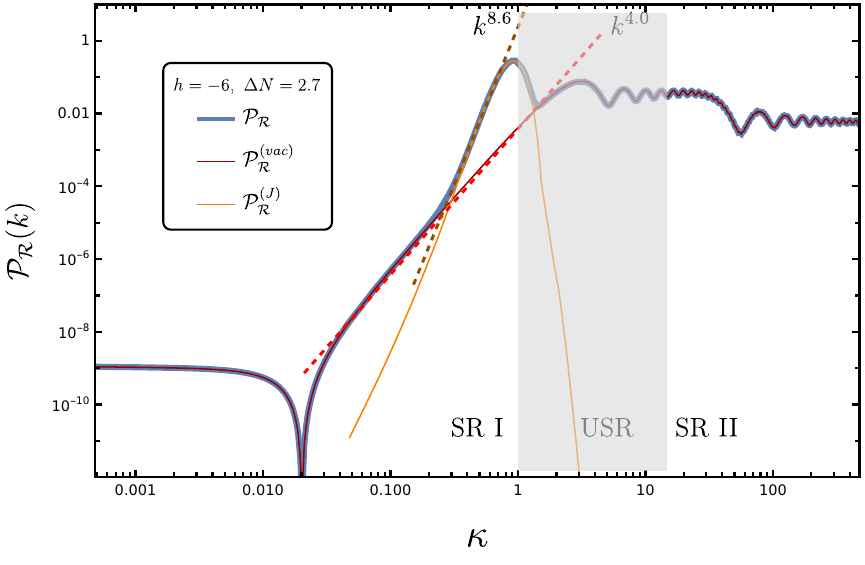}
	\caption{The total power spectrum \eqref{eq:P_total_2_talk} vs  $\kappa \equiv \frac{k}{k_i}$  for the parameters  $\tilde{\alpha}=25$, $h=-6$ and $\Delta N=2.7$. There are two pronounced peaks: the first peak at $\kappa=1$ is due to gauge field production while the 
	second peak at $\kappa>1$ corresponds to standard USR enhancement. 
The scaling of the power spectrum is different for the vacuum and sourced fluctuations. After the dip, there is the universal scaling ${\cal P}_\calR \propto k^4$ which is a well-known feature of USR phase followed by  the scaling $ {\cal P}_\calR \propto k^{8.6}$ just prior to the first peak. }
	\label{fig:power}
\end{figure}

\begin{figure}[t]
	\centering
	\includegraphics[width=0.5\linewidth]{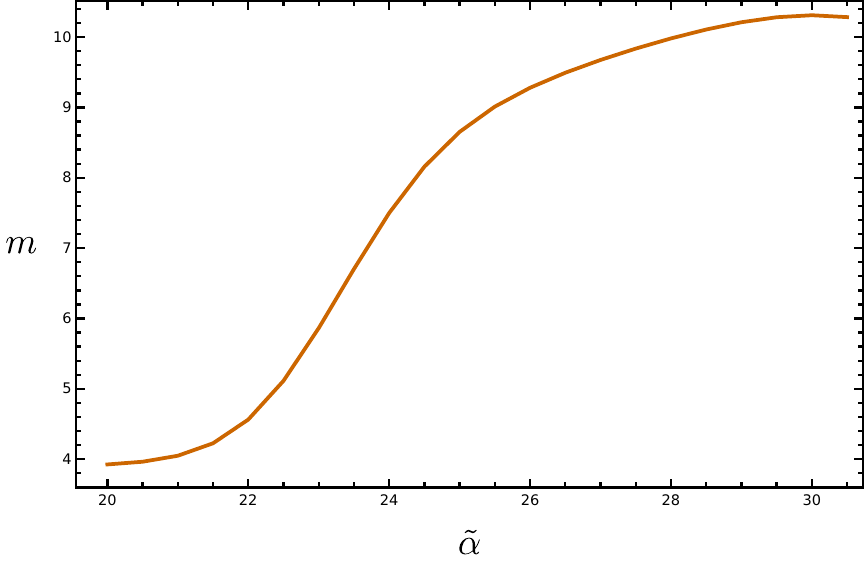}
	\caption{The scaling parameter $m$ defined via ${\cal P}_{\cal R} \propto k^m$ for the  power spectrum  prior to the first peak in terms of the coupling $\tilde{\alpha}$.  For $\tilde{\alpha} > 27$, the power spectrum exceeds unity and the perturbative treatment breaks down. }
	\label{fig:alpha_m}
	\vspace{0.5cm}
\end{figure}

The behaviour of the total power spectrum \eqref{eq:P_total_2_talk} is plotted in Fig. \ref{fig:power}. To generate this plot,  we consider a model with the coupling $\tilde{\alpha}=25$ and the instability parameter $\xi$  obtained as  in  Fig. \ref{fig:xi}. There are a number of interesting features which can be seen in this figure:

\begin{itemize}

	\item There is the plateau on large scales associated to the modes which leave the horizon during the SR1 phase. The amplitude of the power spectrum for these perturbations is fixed by the COBE normalization on $k_{\rm CMB} = 0.05 \,{\rm Mpc}^{-1}$ with ${\cal P}_{_{\rm CMB}} \simeq 2.1 \times 10^{-9}$. 
	
	\item The USR phase starts when the mode $k_i \simeq 2 \times 10^9\,{\rm Mpc}^{-1}$ leaves the horizon which we scale to $\kappa \equiv k/k_i=1$. There is a peak in power due to the source term $\calP^{(J)}_\calR$ \eqref{P_J_2} induced by the inverse decay effect.  The amplitude of this peak depends on the coupling $\tilde{\alpha}$. For the model under consideration, the peak disappears when $\tilde{\alpha} \leq 20$ and diverges (${\cal P}_{\cal R} \gtrsim 1$) for $\tilde{\alpha} > 27$. However, it is important that this peak, if exists, is located at $\kappa=1$ as the effects of gauge field reaches its maximum just prior to the USR phase where the instability parameter 
$\xi$ reaches its maximum value. 
	
\item Prior to the USR phase there is a dip in power spectrum at $k_{\rm d}< k_i$ where \cite{Firouzjahi:2023lzg}
\begin{align}
\label{k_d}
-k_{\rm d}\tau_i \simeq \sqrt{\frac{5h}{4(h-6)}} e^{-\frac{3}{2}\Delta N} \, .
\end{align}
The dip is followed by a universal scaling
${\cal P}_{\cal R} \propto k^4$, a
phenomenon which was observed previously in \cite{Byrnes:2018txb,Cole:2022xqc,Carrilho:2019oqg,Ozsoy:2021pws,Pi:2022zxs} 
as well, see also \cite{Briaud:2025hra} for detail studies of the properties of the dip and peak of USR power spectrum.  In our setup,  just prior to the first peak, 
this universal scaling is followed by ${\cal P}_{\cal R} \propto k^{m}$  with $m >4$ as the contributions from the gauge field dominate over the vacuum fluctuations during a limited interval with the index $m$ depending on the coupling $\tilde{\alpha}$. For example, for the value  $\tilde{\alpha}= 25$ considered in Fig. \ref{fig:power}, we have $m\simeq 8.6$.  In Fig. \ref{fig:alpha_m}, we have plotted the scaling parameter $m$ vs $\tilde{\alpha}$. As seen, the larger is the coupling $\tilde{\alpha}$, the larger is the scaling parameter $m$. 
	
\item After the  peak induced by the gauge field particle creations,  there is a second peak in power spectrum for $\kappa>1$ which is the standard enhancement in power spectrum due to USR mechanism \cite{Firouzjahi:2023lzg}. There are oscillatory features superimposed on a plateau after the USR peak, representing the modes which are deep inside the horizon when  the 
USR phase starts.

\end{itemize}

\begin{figure}[t!]
	\centering
	\includegraphics[width=0.8\linewidth]{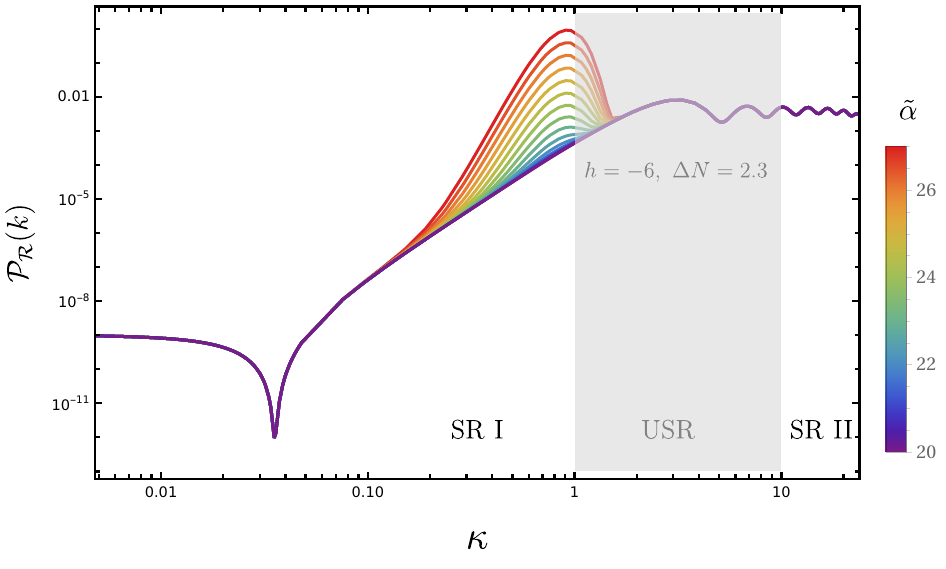}
	\caption{The total power spectrum \eqref{eq:P_total_2_talk} in terms of $\kappa $ for various coupling $\tilde{\alpha}$. For large  $\tilde{\alpha} $, the contribution of the gauge field  $\calR_{\bf k}^{(J)}$ may actually \emph{dominate} over the contribution of the usual vacuum fluctuations $\calR_{\bf k}^{(\mathrm{vac})}$.  In our setup, the first peak associated to $\calR_{\bf k}^{(J)}$ {disappears} for $\tilde{\alpha} <20$, while exhibiting divergent behaviour (${\cal P}_\calR \gtrsim 1$) when $\tilde{\alpha}> 27$, indicating the breakdown of the perturbative regime. }
	\label{fig:power_alpha}
	\vspace{1cm}
\end{figure}

In Fig. \ref{fig:power_alpha}, we have plotted the total power spectrum for various coupling $20< \tilde{\alpha} < 27$. As seen, the contribution  from the gauge field particle productions  $\calP^{(J)}_\calR$ on the power spectrum disappears for $\tilde{\alpha} \leq 20$. For $\tilde{\alpha} > 27$, the perturbative regime breaks down and the power spectrum exceeds unity. An interesting feature is that the
first peak induced by the gauge field (if it exists) can be larger than the second peak induced by the standard USR  contribution.  This double peak feature can have interesting implications for PBHs formation and induced GWs production \cite{Firouzjahi:2023lzg}. \\

\begin{figure}[t!]
	\centering
	\includegraphics[width=0.49\linewidth]{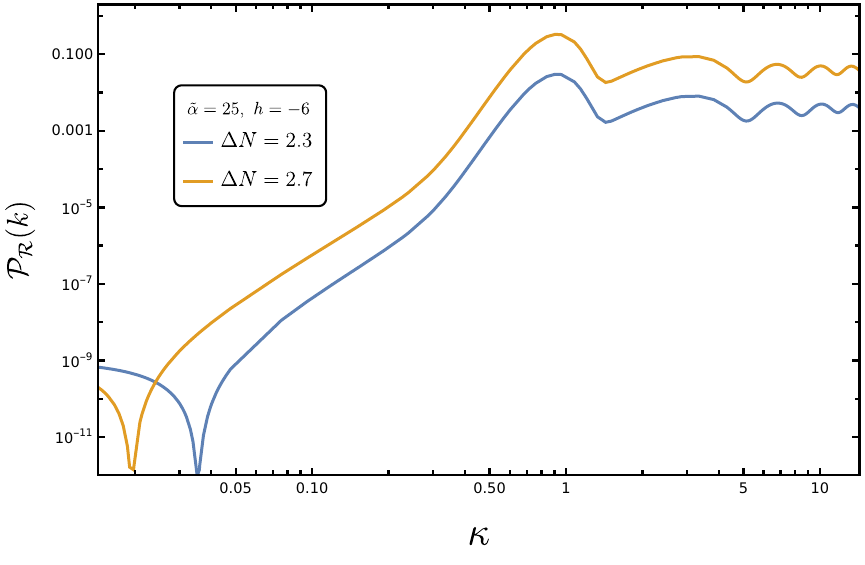}
	\includegraphics[width=0.49\linewidth]{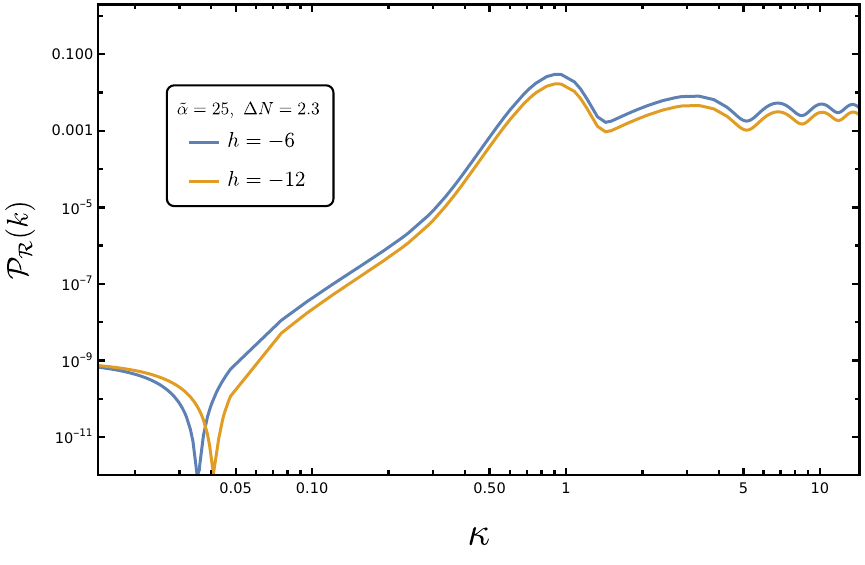}
	\caption{The total power spectrum \eqref{eq:P_total_2_talk} for different values of $h$ and $\Delta N$ while the dimensionless coupling $\tilde{\alpha} =25$ is held fixed. The amplitudes of peaks are sensitive to the duration of the USR phase $\Delta N$ while they are less sensitive to the sharpness parameter $h$. }
	\label{fig:power_h_deltaN}
\end{figure}

In Fig. \ref{fig:power_h_deltaN}, we have shown how  changing the parameters  of the USR phase can affect the total power spectrum 
while keeping the coupling $\tilde{\alpha}$  fixed. As seen, the amplitude of the first peak  is  affected not only by the coupling $\tilde{\alpha}$ but also by the parameters $h$ and $\Delta N$. For example, by increasing  the duration of the USR phase $\Delta N$, the amplitudes of both  peaks increase. \\

In order to remain in perturbative regime ${\cal P}_{\calR} <1$,  we need $\tilde{\alpha} < 27$ in our model. Larger values of $\alpha$  cause overproduction of gauge field particles which in turn induce large 
corrections  in curvature perturbation power spectrum.  
The maximum value  of the instability parameter allowed is  $\xi_i \simeq 4.3$ as shown in Fig. \ref{fig:xi}.

\section{Bispectrum}
\label{sub:3point}

We now turn to the calculation of the bispectrum, the three-point correlation function  of  curvature perturbations. In the Fourier space,  the three-point correlation function is given by,
\begin{align}
	\langle 
	\calR_{\bf k_1}\, 
	\calR_{\bf k_2}\, 
	\calR_{\bf k_3} 
	\rangle
	\equiv 
	{\cal B}_{\calR}(k_1,k_2,k_3) ~\delta^{(3)}(\bfk_1+\bfk_2+\bfk_3)
\end{align}
in which $k_i = \abs{\bf k_i}$ and ${\cal B}_{\calR}$ is called the bispectrum which depends only on the magnitudes of the three external momenta. The factor $\delta^{(3)}(\bfk_1+\bfk_2+\bfk_3)$ represents the translation invariance associated to the homogeneity of the background.

The bispectrum ${\cal B}_{\calR}$ can be parameterized in terms of the non-linearity  parameter $f_{_{\rm NL}}$, which is a measure of  the amplitude of non-Gaussianity  \cite{Barnaby:2011vw}, 
\begin{align}\label{BR}
	{\cal B}_{\mathcal R}\left(k_{1},k_{2},k_{3}\right) \equiv 
	\dfrac{3}{10} \, \left( 2 \pi \right)^{5/2} \, f_{_{\rm NL}}(k_1, k_2, k_3) \, \mathcal{P}_{\cal R}^2 ~ \dfrac{\sum_{i=1}^{3} k_{i}^3}{
			\prod_{n=1}^{3} k_{i}^3} \,.
\end{align}
Usually, the non-Gaussianity parameter $ f_{_{\rm NL}}$ is best suited for the local-shape non-Gaussianity in which the non-linear curvature perturbation in real space  is expanded as follows, 
\begin{equation}
	\label{eq:fNL}
	{\calR} \left( {\bf x} \right) = \calR_g \left( {\bf x} \right) + \frac{3}{5} \, f_{_{\rm NL}} \,\left[ \calR_g^2 \left( {\bf x} \right) - \langle  \calR_g^2 \left( {\bf x} \right) \rangle \right] \,,
\end{equation}
where $\calR_g$ is a Gaussian field. However, we  keep using the ansatz (\ref{BR}) as a useful measure of the amplitude and the shape of non-Gaussianity. 

As we are interested in both the amplitude and the shape of bispectrum, we 
rewrite Eq. (\ref{BR}) as follows, 
\begin{equation}
	\langle \calR_{\bf k_1} \,  \calR_{\bf k_2} \,  \calR_{\bf k_3} \rangle = \frac{ 3 }{ 10 } \, \left( 2 \pi \right)^{5/2} \, f_{_{\rm NL}} ~ \calP_\calR \left( k \right)^2 ~
	\frac{\delta^{(3)} \left( {\bf k_1} +  {\bf k_2} +  {\bf k_3}  \right)}{k^6} \, \frac{1+x_2^3+x_3^3}{x_2^3 \, x_3^3} \, ,
	\label{local_bispectrum}
\end{equation}
where the momenta $k_i$ are scaled in the following way,  
\begin{equation}
	\label{notation_x}
	\vert {\bf k_1} \vert  \equiv k \;\;,\;\;  \vert {\bf k_2} \vert  \equiv x_2 \,  k \;\;,\;\; \vert {\bf k_3} \vert \equiv x_3 \, k \,.
\end{equation}

Since the source term $\calR_k^{(J)}$ adds incoherently with the vacuum perturbations $\calR_k^{(\mathrm{vac})}$ with different creation and annihilation operators, the three-point function is divided into two separate contributions,
\begin{align}
	\langle \calR_{\bf k_1} \,  \calR_{\bf k_2} \,  \calR_{\bf k_3} \rangle &=\langle \calR_{\bf k_1}^{(\mathrm{vac})} \,  \calR_{\bf k_2}^{(\mathrm{vac})} \,  \calR_{\bf k_3}^{(\mathrm{vac})} \rangle + \langle \calR_{\bf k_1}^{(J)} \,  \calR_{\bf k_2}^{(J)} \,  \calR_{\bf k_3}^{(J)} \rangle \,.
\end{align}
The first term above represents the contribution from the usual inflaton perturbations in USR setup. Its amplitude and shape are studied in the previous works, see for example \cite{Cai:2018dkf, Firouzjahi:2023aum, Pi:2022ysn, Cai:2022erk, Firouzjahi:2023xke, Namjoo:2024ufv}. In particular,
for the modes which leave the horizon during the USR phase, we have a local-shape non-Gaussianity with the amplitude \cite{Cai:2018dkf, Firouzjahi:2023aum}
\ba
\label{fNL-USR}
f_{_{\rm NL}}^{\mathrm{USR}} = \frac{5 h^2 }{(h-6)^2} \, .
\ea

Here we are mainly interested in the bispectrum induced by the gauge field via the source term  $\calR_{\bf k}^{(J)} $.  Using Eq. \eqref{R_J_00} for the induced curvature perturbations {$\calR_{\bf k_i}^{(J)}$},  we obtain
\begin{eqnarray}
	\langle 
	\calR_{\bf k_1}^{(J)} \, 
	\calR_{\bf k_2}^{(J)} \, 
	{\calR_{\bf k_3}^{(J)}}
	\rangle 
	&=& \left(\frac{-\tilde{\alpha}}{M_{\rm Pl}}\right)^3
	\frac{H^6 e^{6\pi\xi_k}}{2^3k^{12}M_{\rm Pl}^3\sqrt{8\epsilon_f^3}} \,  \delta^{(3)} \left( {\bf k_1} + 
	{\bf k_2} +  {\bf k_3} \right) \int \frac{d^3 q {\, d^3 q_2 \, d^3 q_3}}{\left( 2 \pi \right)^{9/2}} \times 
	\Big[
	\varepsilon \, {\rm products}
	\Big]
	\nonumber\\
	&&\hspace{.5cm}\times
	\abs{\bfk_1-\vb{q}}^{1/4}\abs{\vb{q}}^{1/4} \left(
	\abs{\bfk_1-\vb{q}}^{1/2}+\abs{\vb{q}}^{1/2}
	\right) ~	{\cal I}(\xi_k,\abs{\bfk_1},|\vb{q}-{\bfk_1}|)
	\nonumber\\
	&&\hspace{.5cm}\times 
  	\abs{\bfk_2-\vb{q}_2}^{1/4}\abs{\vb{q}_2}^{1/4} \left(
  	\abs{\bfk_2-\vb{q}_2}^{1/2}+\abs{\vb{q}_2}^{1/2}
  	\right) ~	{\cal I}(\xi_k,\abs{\bfk_2},|\vb{q}_2-{\bfk_2}|)
  	\nonumber\\
  	&&\hspace{.5cm}\times
  	\abs{\bfk_3-\vb{q}_3}^{1/4}\abs{\vb{q}_3}^{1/4} \left(
  	\abs{\bfk_3-\vb{q}_3}^{1/2}+\abs{\vb{q}_3}^{1/2}
  	\right) ~	{\cal I}(\xi_k,\abs{\bfk_3},|\vb{q}_3-{\bfk_3}|) \nonumber\\
  	&&\hspace{.5cm}\times
  	\delta^{(3)}(\bf{k}_1-\bf{q}+\bf{q}_2) \, \delta^{(3)}(\bf{q}+\bf{k}_3-\bf{q}_3) \,,
	\label{R3}
\end{eqnarray}
{where ${\cal I}$ is the integral defined via Eq. \eqref{eq:calI}. In addition,} to simplify the analysis we have assumed that $\xi_k$ is the same for $k_1, k_2$ and $k_3$ and used the following relations,
\begin{align}
	\langle \vec{\mathcal{O}}_{\bf{q}_1} \cdot \vec{\mathcal{O}}_{\bf{k}_1-\bf{q}_1} ~~ \vec{\mathcal{O}}_{\bf{q}_2} \cdot \vec{\mathcal{O}}_{\bf{k}_2-\bf{q}_2}~~ \vec{\mathcal{O}}_{\bf{q}_3} \cdot \vec{\mathcal{O}}_{\bf{k}_3-\bf{q}_3} \rangle = 8 \Big[
	\varepsilon \, {\rm products}
	\Big] \, \delta^{(3)} \left( {\bf k_1} + 
	{\bf k_2} +  {\bf k_3} \right)
	\nonumber\\
	\times
	\delta^{(3)}(\bf{k}_1-\bf{q}_1+\bf{q}_2) \, \delta^{(3)}(\bf{q}_1+\bf{k}_3-\bf{q}_3) \,,
\end{align}
and
\begin{align}
	\Big[
	\varepsilon \, {\rm products}
	\Big]  \equiv \left[ \vec{\epsilon} \left( {\bf q} \right) \cdot \vec{\epsilon} \left( {\bf k_1} - {\bf q} \right) \right] \,
	\left[ \vec{\epsilon} \left( {\bf q} - {\bf k_1} \right) \cdot \vec{\epsilon} \left(  - {\bf q} - {\bf k_3}  \right) \right] \,
	\left[ \vec{\epsilon} \left( {\bf q} + {\bf k_3} \right) \cdot \vec{\epsilon} \left(   - {\bf q} \right) \right] 
	 \,.
\end{align}

The three-point function \eqref{R3} is determined both by the amplitude and the shape of the triangle defined by the vectors ${\bf k_i}$. Using the 
scaling \eqref{notation_x}, we reparameterize the three-point function \eqref{R3} as,
\begin{eqnarray}
	\big\langle 
	\calR_{\bf k_1}^{(J)} \, 
	\calR_{\bf k_2}^{(J)} \, 
	\calR_{\bf k_3}^{(J)} 
	\big\rangle 
	&=& \frac{3}{10} \left( 2 \pi \right)^{5/2} \, f_3 \left( \xi_k ;\, x_2 ,\, x_3 \right)   \Big(\dfrac{6}{\abs{h}}  {\cal P}_{_{\rm CMB}} \, e^{3\Delta N+2 \pi \xi_k} \Big)^3  \,   \nonumber\\ 
	&&\hspace{.1cm}\times
	\frac{\delta^{(3)} \left( {\bf k_1} +  {\bf k_2} +  {\bf k_3}  \right)}{k^6} \, \frac{1+x_2^3+x_3^3}{x_2^3 \, x_3^3} \,, 
	\label{RJ3}
\end{eqnarray}
where the function  $f_3 \left( \xi_k ;\, x_2 ,\, x_3 \right)$ is defined via, 
\begin{align}
	f_3 \left( \xi_k ;\, x_2 ,\, x_3 \right) &= \frac{5}{3\pi} \, \frac{\xi_{_{\rm CMB}}^3}{x_2 \, x_3 \left[ 1 + x_2^3 + x_3^3 \right]}
	\nonumber\\
	\hspace{.1cm}
	&\times
	\int d^3 r 
	~ {\cal G}\big({\bf r},{\bf r} - {\hat \bfk}_1,1 \big) 
	~ {\cal G}\big({\bf r} - {\hat \bfk}_1,{\bf r} +x_3 {\hat \bfk}_3,x_2 \big)
	~ {\cal G}\big({\bf r} +x_3 {\hat \bfk}_3,{\bf r},x_3 \big) \,,
	\label{f3}
\end{align}
with the function ${\cal G}$ defined as, 
\begin{align}
	{\cal G}\left({\bfq},{\bf p},x \right) \equiv \abs{{\bfq}}^{1/4}\abs{{\bf p}}^{1/4} \left(
	\abs{{\bfq}}^{1/2}+\abs{{\bf p}}^{1/2}
	\right) ~ \vec{\epsilon} \left( {\bf q} \right) \cdot \vec{\epsilon}^{\,\, *} \left( {\bf p} \right) ~	{\cal I}\Big(\dfrac{\xi}{x},\abs{\bfq},\abs{{\bf p}}\Big) \,.
	\label{cal_G}
\end{align}

From the above expressions, we define the ``effective'' 
nonlinearity parameter $f_{_{\rm NL}}^{\mathrm{eff}}$ generated by the gauge field particle production. Comparing Eq. \eqref{local_bispectrum} with \eqref{RJ3}, we obtain the following relation between 
$f_{_{\rm NL}}^{\mathrm{eff}}$ and $ f_3 \left( \xi_k ;\, x_2 ,\, x_3 \right)$, 
\begin{equation}
	f_{_{\rm NL}}^{\mathrm{eff}}(\xi_k;x_2,x_3) = \frac{\Big(\dfrac{6}{\abs{h}}  {\cal P}_{_{\rm CMB}} \, e^{3\Delta N+2 \pi \xi_k} \Big)^3}{{\cal P}_\calR \left(k \right)^2} \, f_3 \left( \xi_k ;\, x_2 ,\, x_3 \right)
	\label{fNL_eff} \,.
\end{equation}

\begin{figure}
	\centerline{
		\includegraphics[angle=0,width=0.8\textwidth]{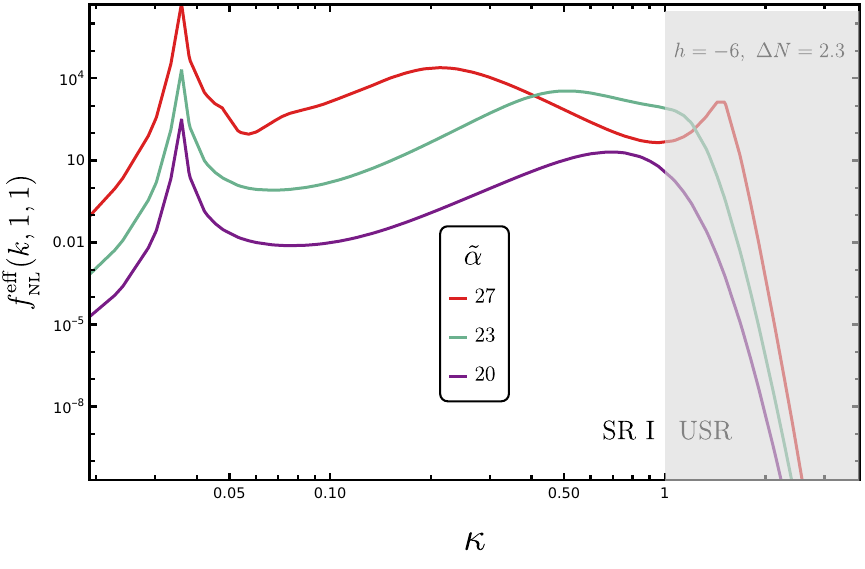}
	}
	\caption{The effective nonlinear parameter $f_{_{\rm NL}}^{\rm eff}$ \eqref{fNL_eff} for the equilateral shape $x_2=x_3=1$ in terms of 
	$\kappa$. The location, shape, and number of peaks vary for different values of $\tilde{\alpha}$. The beginning of USR corresponds to $\kappa=1$ after which $f_{_{\rm NL}}^{\rm eff}$ falls off quickly.  
\vspace{1cm}
	}
	\label{fig:f3_1}
\end{figure}

\begin{figure}
	\centerline{
		\includegraphics[angle=0,width=0.8\textwidth]{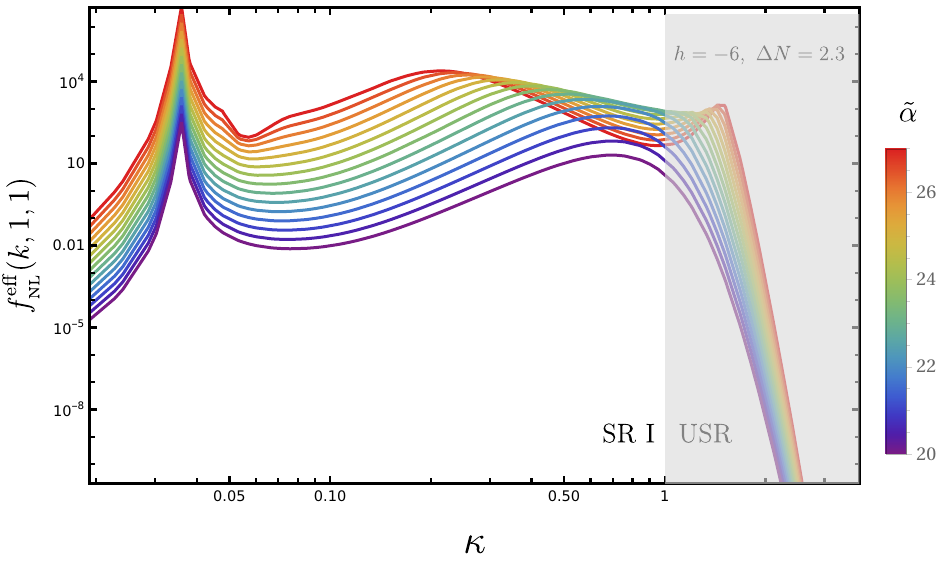}
	}
	\caption{The effective nonlinear parameter $f_{_{\rm NL}}^{\rm eff}$ \eqref{fNL_eff} for the equilateral shape $x_2=x_3=1$  for various coupling parameter $\tilde{\alpha}$.  This is the same plot as in Fig. \ref{fig:f3_1} but with extended values of  $\tilde{\alpha}$.  There are a number of features which are discussed in the text. }
	\label{fig:f3}
\end{figure}

In Figs. \ref{fig:f3_1} and \ref{fig:f3} we have plotted $f_{_{\rm NL}}^{\mathrm{eff}}$ generated from the gauge field production  in the equilateral shape where $x_2=x_3=1$. We have fixed the USR parameters $h$ and $\Delta N$ while the coupling $\tilde{\alpha}$ is varied in the range $20 < \tilde{\alpha} < 27$. There are a number of features in {these figures:}

\begin{itemize}
	\item The instability parameter $\xi_k$ in \eqref{fNL_eff} represents the value of  $\xi$ when the mode $k$ leaves the horizon. 
Since the relation between $\xi$ and $k$ is one-to-one, we have plotted $f_{_{\rm NL}}^{\mathrm{eff}}(k;1,1)$.
	
\item There is a peak in $f_{_{\rm NL}}^{\mathrm{eff}}$  which  coincides with the location of dip in the power spectrum at $k_{\rm d}$ \eqref{k_d}. This is because the factor ${\cal P}_\calR \left(k \right)^2$ appears in the denominator of Eq. \eqref{fNL_eff}  which has a dip at $k_{\rm d}$.
	
\item As the coupling $\tilde{\alpha}$ increases, the nonlinearity parameter becomes larger and its second peak  shifts to the smaller values of $k$. As a result, the peak of bispectrum does not coincide with the peak in power spectrum, a feature which was noticed in USR setup 
in \cite{Namjoo:2024ufv} as well.

\item For $\tilde{\alpha} > 25$, there are two additional peaks in the bispectrum located on  both sides of the mode $k_i$ (or $\kappa=1$).   For this range of $\tilde{\alpha}$, the local minimum of bispectrum is close to the peak of the power spectrum.

\item As can be seen in Figs. \ref{fig:f3_1} and \ref{fig:f3}, the amplitude of  $f_{_{\rm NL}}^{\mathrm{eff}}$ falls off rapidly during and after the USR phase. This is because during the USR phase the instability parameter $\xi$ becomes negligible and the tachyonic process to generate gauge field particles becomes inefficient. 
\end{itemize}

In the above analysis  we have  calculated the bispectrum of curvature perturbations generated via the gauge field particle production. To obtain the 
total bispectrum, the contribution from the usual inflaton vacuum perturbation 
should be included as well. As these two contributions are uncorrelated, we can write, 
\begin{align}
	f^{^{\rm tot}}_{_{\rm NL}} = f^{^{\rm USR}}_{_{\rm NL}} 
	+ f_{_{\rm NL}}^{\mathrm{eff}} \, .
\end{align}
While the USR contribution $f^{^{\rm USR}}_{_{\rm NL}} $ has mostly the local shape, but the contribution from the gauge field production has a complicated shape.
In our numerical analysis above we have only presented the equilateral shape for $f_{_{\rm NL}}^{\mathrm{eff}}$ as generating the plots for other shapes is numerically challenging. One difficulty in dealing with other shapes is that the parameter $\xi_k$ takes different values for different values of $k_i$ which makes the corresponding integrals complicated to be calculated numerically.

\section{Summary and Discussions}
\label{sec:sum}

In this work we have studied a model of inflation in which  an axion field plays the role of the inflation field. We considered the potential with an inflection point which supports an intermediate USR phase during inflation. In order to be consistent with cosmological observations, the early stage is in the form of SR inflation during which the large CMB scales leave the horizon. The intermediate USR phase is extended for a limited  number of e-folds followed by an extend phase of secondary SR inflation and reheating. This 
three-phase picture SR-USR-SR  has been used extensively in recent literature to generate PBHs and GWs. 

As in standard axion inflation models, the Chern-Simons interaction \eqref{int} 
causes  one polarization of the gauge field perturbations to become 
tachyonic, yielding to non-perturbative gauge field particle production.  As is well-known, the generated gauge field particles backreact on the inflaton perturbations via the inverse decay process. As a result, the curvature perturbations has two parts: the usual inflationary vacuum contribution and the contribution from the gauge field particles via the inverse decay effect.  
The efficiency of the inverse decay process depends largely on the instability parameter $\xi\propto \sqrt\epsilon$. In conventional models of axion inflation in SR setups,  $\epsilon$ increases slowly towards the end of inflation 
and correspondingly $\xi$ increases slowly as well. This in turn causes the small scales at the end of inflation to experience the gauge field particles production. However, a novel feature in our setup is that since $\epsilon$ falls off exponentially during the USR phase, then the gauge field particles production and the inverse decay effects shut off efficiently during and after the USR phase.  This is a key property allowing us 
to engineer the onset of the USR phase to terminate the gauge field particles production and prevent 
$\xi$ to enter the strong coupling regime. There are a number of interesting effects in this setup which we summarize below. 

The power spectrum induced by the gauge field field 
$\calP^{(J)}_\calR (k)$ reaches its maximum at the start of the USR phase. This is because the maximum value of $\xi$ occurs at the end of SR1 phase when the USR phase takes over. On the other hand, the power spectrum for the modes which leave the horizon during the USR phase experiences  the usual USR enhancement. Correspondingly, there are two distinct sources of power enhancement in our setup: (a) the enhancement in power spectrum from the inverse decay effect, and (b): the enhancement by the USR mechanism. The combination of these two effects appear non-trivially in the total power spectrum which were presented in Figs.  \ref{fig:power} and \ref{fig:power_alpha}. As can be seen in these figures, there is a dip for the modes which leave the horizon before the USR phase starts, followed by the universal $k^4$ scaling as common in other USR setups. However, an 
interesting  feature in our model is that the power spectrum develops a non-trivial scale dependence prior to the peak in the form $\calP_\calR \propto k^m$. The index $m$ depends on the gauge coupling $\tilde \alpha$ as can be seen in 
Fig. \ref{fig:alpha_m}.   For example, in our setup this can reach to $m \simeq 8.6$ for $\tilde \alpha = 27$. For larger values of $\tilde \alpha $, the power spectrum becomes larger than unity, indicating the  breakdown of the perturbative treatment. Another feature of this setup is that the power spectrum can have two peaks, the first peak being associated to the inverse decay effect appearing at $\kappa=1$ while the second peak is the usual USR peak which happens for $\kappa >1$. Furthermore, for large values of $\tilde \alpha$, the first peak can be larger than the USR peak as can be seen in Fig. \ref{fig:power_alpha}. 

We have calculated the bispectrum in this setup. The non-Gaussianity associated to the USR dynamics has largely the local shape with the amplitude of $f_{_{\rm NL}}^{\mathrm{USR}}$ given in Eq. (\ref{fNL-USR}). However, the  contributions of the gauge field corrections in curvature perturbation induces non-trivial shapes of bispectrum. While we have provided the general expression for the bispectrum in Eq. (\ref{fNL_eff}) but we only presented the numerical plots for the equilateral shape as calculating the convolution integrals such as in Eq. (\ref{f3}) 
for other shapes becomes numerically expensive.  Depending on the value of the coupling $\tilde \alpha$, the amplitude of non-Gaussianity  $f_{_{\rm NL}}^{\rm eff}$ induced from the gauge field production can have  non-trivial form and multiple peaks.  

As the power spectrum shows non-trivial scale-dependence and peaks, it can have interesting implications for PBHs formation and GWs. 
Specifically, if the coupling $\tilde \alpha$ is large  enough so the first peak is pronounced, then the mechanism of the PBHs formation should be considered 
carefully. On the other hand, for the GWs generation we have two non-trivial sources. First, we have the induced GWs from the scalar sector as studied for example in \cite{Firouzjahi:2023lzg}. Second, we have the GWs generated from the gauge field sector \cite{Cook:2011hg,Barnaby:2011qe,Domcke:2016bkh, Sorbo:2011rz,Barnaby:2012xt,Namba:2015gja, Garcia-Bellido:2016dkw, Obata:2016xcr,Dimastrogiovanni:2016fuu, Dimastrogiovanni:2018xnn, Gorji:2020vnh, Salehian:2020dsf,  Talebian:2022cwk}. We would like to come back to these interesting implications in future studies. 

In this work, in order to perform the calculations analytically, we have considered an idealized situation in which the transitions from  SR1 to USR and then to the second SR2 phase take place instantaneously. In addition, we have assumed that the transition from the USR phase to SR2 phase is sharp such that the system reaches its attractor phase in short amount of time. Both of these are idealized assumptions which can be relaxed in a more realistic situation but this requires a full numerical analysis as the calculations can not be tracked analytically.

\vspace{1cm}

\textbf{Acknowledgments:} We would like to thank T. Rostami for collaboration at the early stage of this work.  H. F. and A. T. are partially supported by INSF of Iran under the grant numbers  4022911 and 4038049.

%
\small
\bibliography{ref}
\bibliographystyle{JHEP}

\end{document}